\documentclass[12pt,a4paper]{article}
\usepackage{amsmath}
\usepackage{amsthm}
\usepackage{amsfonts}          % if you want the fonts
\usepackage{amssymb}           % if you want extra symbols

\usepackage[all]{xy}

\usepackage{color} 
\usepackage{graphics}
\usepackage{graphicx}

\newcommand{\diff}{{\rm d}}
\newcommand{\dslash}{\mbox{$\partial$ \kern-.92em  \big /}}
\newcommand{\transverse}{\ensuremath{\,\mbox{\lower 0.02ex\hbox{\small$\cap$} 
    \kern-1.03em $\top$}\,}}

\newcommand{\Id}[1]{\mathop{\rm id}_{#1}\nolimits}

\newcommand{\coker}{{\mathop{\rm coker}\nolimits\,}}

\newcommand{\eqdef}{\stackrel{\rm def}{=}}

\newcommand{\bra}[1]{\ensuremath{\left< #1 \right|}}
\newcommand{\ket}[1]{\ensuremath{\left| #1 \right>}}

\newcommand{\R}[1]{\ensuremath{{\mathbb{R}^{#1}}}}
\newcommand{\C}[1]{\ensuremath{{\mathbb{C}^{#1}}}}
\newcommand{\Hbb}[1]{\ensuremath{{\mathbb{H}^{#1}}}}

\newcommand{\Ktheory}{K--theory}
\newcommand{\Ktheories}{K--theories}
\newcommand{\Kgroup}{K--group}
\newcommand{\Kgroups}{K--groups}
\newcommand{\KRtheory}{KR--theory}
\newcommand{\KRtheories}{KR--theories}

\newcommand{\KRgroups}{KR--groups}

\newcommand{\Tor}{{\mathop{\rm Tor}\nolimits}}

\newcommand{\Mat}{\mathop{\rm Mat}\nolimits}
\newcommand{\End}{{\mathop{\rm End}\nolimits}}

\newcommand{\GL}{\mathop{GL}\nolimits}

\newcommand{\CliffGen}[2]{{\mathop{\mathbf{C}}\nolimits_{#1}^{#2}}}
\newcommand{\Cliff}[1]{\CliffGen{}{}{#1}}
\newcommand{\CliffR}[1]{\CliffGen{\R{}}{#1}}

\newcommand{\Dbranes}{D--branes}

\newcommand{\im}{\mathop{\rm Im}\nolimits}
\newcommand{\Tr}{\mathop{\rm Tr}\nolimits}
\newcommand{\partBox}[2]{\hbox{\lower -0.1cm
  \hbox{$\displaystyle {\scriptstyle {#1}}
    \mathop{\fbox{\phantom{e}}}_{#2}$}}
}

\newenvironment{displayarray}%
{\everymath{\displaystyle\everymath{}}\array}%
{\endarray}

\newtheorem{theorem}{Theorem}

\newtheorem{definition}{Definition}

\usepackage[active]{srcltx}
\newcommand{\Uone}{U_1}
\newcommand{\Ut}{\widetilde{U}_1}
\newcommand{\Rt}{\widetilde{\Rop}}
\newcommand{\Zt}{\widetilde{\Zop}}
\makeatletter
\let\bra\@undefined
\let\ket\@undefined
\makeatother

\newcommand{\refeq}[1]{(\ref{#1})}

\def\Tr{{\rm Tr}}

\def\I{{\cal I}}

\def\R{{\bf R}}
\def\C{{\bf C}}

\def\RR{R-R }
\def\NSNS{NS-NS }

\def\db{$\bar{\mbox{D}}$}

\def\Zop{\bbbz} 
\def\Rop{\bbbr}   

\def\pmb#1{\setbox0=\hbox{#1}
 \kern-.025em\copy0\kern-\wd0
 \kern.05em\copy0\kern-\wd0
 \kern-.025em\raise.0433em\box0 }

\def\db{$\bar{\mbox{D}}$}
\def\3{\ss}
\def\sq{\hbox{\rlap{$\sqcap$}$\sqcup$}}
\def\qed{\ifmmode\sq\else{\unskip\nobreak\hfil
\penalty50\hskip1em\null\nobreak\hfil\sq
\parfillskip=0pt\finalhyphendemerits=0\endgraf}\fi}

\def\bbbz {{\sf Z\!\!Z}}

\def\bbbr {{\rm I\!R}}

\newcommand{\ket}[1]{|#1\rangle}
\newcommand{\bra}[1]{\langle#1|}

\def\mm {\advance\voffset by -1.5cm
\advance\hoffset by -2.1cm
\textwidth=17.3cm
\textheight=20.0cm}

\def\b,#1,#2(#3){\left|B#1,#2\right>_{#3}}

\def\gso#1,#2{\frac{1}{4}(1#1(-1)^F)(1#2(-1)^{{\tilde F}})}

\def\xxx#1           {{\sf hep-th/#1} }

\def\npb#1(#2)#3     {Nucl. Phys. {\bf B#1} (#2) #3 }
\def\rep#1(#2)#3     {Phys. Rept.{\bf #1} (#2) #3 }
\def\pla#1(#2)#3     {Phys. Lett. {\bf #1A} (#2) #3 }   
\def\plb#1(#2)#3     {Phys. Lett. {\bf #1B} (#2) #3 }
\def\prl#1(#2)#3     {Phys. Rev. Lett.{\bf #1} (#2) #3 }
\def\prd#1(#2)#3     {Phys. Rev. {\bf D#1} (#2) #3 }
\def\ap#1(#2)#3      {Ann. Phys. {\bf #1} (#2) #3 }
\def\rmp#1(#2)#3     {Rev. Mod. Phys. {\bf #1} (#2) #3 }
\def\cmp#1(#2)#3     {Comm. Math. Phys. {\bf #1} (#2) #3 }
\def\mpla#1(#2)#3    {Mod. Phys. Lett. {\bf A#1} (#2) #3 }
\def\ijmp#1(#2)#3    {Int. J. Mod. Phys. {\bf A#1} (#2) #3 }
\def\cqg#1(#2)#3     {Class. Quant. Grav. {\bf #1} (#2) #3 }
\def\am#1(#2)#3      {Adv. Math. {\bf #1} (#2) #3 }
\def\im#1(#2)#3      {Invent. Math. {\bf #1} (#2) #3 }
\def\jhep#1(#2)#3    {JHEP {\bf #1}(#2) #3 }
\def\npps#1(#2)#3    {Nucl.Phys.Proc.Suppl. {\bf #1}(#2) #3 }
\def\jgp#1(#2)#3     {J. Geom. Phys. {\bf #1}(#2) #3 }

\def\dstyler#1       {\documentstyle{report}[#1]}
\def\dstylea#1       {\documentstyle{article}[#1]}
\def\bd              {\begin{document}}
\def\ed              {\end{document}}
\def\be	             {\begin{equation}}
\def\ee              {\end{equation}}
\def\ba              {\begin{eqnarray}}
\def\ea              {\end{eqnarray}}
\def\ni              {\noindent}
\def\bb#1            {\begin{thebibliography}{#1}}
\def\eb              {\end{thebibliography}}

\def\etal {{\em et al.} }
\def\w    {\;_\wedge}
\def\ie   {{\em i.e.}}
\def\ibid {{\em ibid.}}
\def\cf   {{\em c.f.} }

\begin{document}

\thispagestyle{empty}
\def\thefootnote{\fnsymbol{footnote}}
\begin{flushright}
  hep-th/0206158 \\
  HU-EP-02/24 \\
  AEI-2002-042
\end{flushright}

\vskip 0.5cm

\begin{center}\LARGE
{\bf Orientifolds and \Ktheory}
\end{center}

\vskip 1.0cm

\begin{center}
{\large V. Braun,\footnote{E-mail address: {\tt volker.braun@physik.hu-berlin.de}}}

\vskip 0.5cm
{\it Humboldt Universit\"at zu Berlin, Institut f\"ur Physik 
\\ Invalidenstrasse 110, D-10115 Berlin, Germany}
\end{center}

\vskip 0.5cm
\begin{center}
{\large B. Stefa\'nski, jr.\footnote{E-mail address: {\tt bogdan@aei.mpg.de}}}
\vskip 0.5cm
{\it Max-Planck-Institut f\"ur Gravitationsphysik, Albert-Einstein 
Institut \\ Am M\"uhlenberg 1, D-14476 Golm, Germany}
\end{center}

\vskip 1.0cm

\begin{center}
May 2002
\end{center}

\vskip 1.0cm

\begin{abstract}
\noindent Recently it has been shown that D-branes in orientifolds are not
always described by equivariant Real \Ktheory. In this paper we define
a previously unstudied twisted version of equivariant Real \Ktheory{}
which gives the D-brane spectrum for such orientifolds. We find that
equivariant Real \Ktheory{} can be twisted by elements of a
generalised group cohomology. This cohomology classifies all
orientifolds just as group cohomology classifies all orbifolds.  As an
example we consider the $\Omega\times\I_4$ orientifolds.  We
completely determine the equivariant orthogonal \Ktheory{}
$KO_{\Zop_2}(\R^{p,q})$ and analyse the twisted versions. Agreement is
found between \Ktheory{} and Boundary Confromal Field Theory (BCFT) 
results for both integrally- and torsion-charged D-branes.
\end{abstract}

\vfill

\setcounter{footnote}{0}
\def\thefootnote{\arabic{footnote}}
\newpage
% ================= end titlepage    ========================

\section{Introduction}

Brane-anti-brane annihilation~\cite{Snbps,Srednicki} is the physical
manifestation of the equivalence relations that define \Ktheory.  Lower
dimensional D-branes can be thought of as non-trivial tachyon bundles
on D\db 9-brane pairs in Type IIB~\cite{WittK} or non-BPS D9-branes in
Type IIA~\cite{HorK}. As such, the spectrum of stable D-branes is
classified by \Ktheory{}~\cite{MM,WittK,HorK}. In this construction
D-branes on orbifolds are described by equivariant \Ktheory{} while
D-branes in Type I and its T-duals correspond to Real 
\Ktheory.\footnote{Elements of \Ktheory{} are pairs of
isomorphism classes of complex bundles on a manifold; in equivariant \Ktheory{}
a group acts on the bundles with the corresponding map on fibres being 
linear; Real \Ktheory{} or \KRtheory{} is similar to equivariant
\Ktheory{} but with an element acting anti-linearly (for example
by complex conjugation) on the fibres.}

It has been known for some time that certain orbifolds admit discrete
torsion~\cite{VafaT}. These are classified by the projective
representations of the orbifold group $G$, or equivalently by group
cohomology\footnote{We use $\Uone$ instead of $U(1)$ for typographical
  reasons.} $H^2(G,\Uone)$. The allowed choices give different closed
string backgrounds and hence also different D-brane spectra. It is
possible to define twisted equivariant \Ktheories{} which describe the
D-brane spectrum in orbifolds with discrete torsion~\cite{WittK} (see
also~\cite{gomis}. We review this construction section~\ref{sec:TwistedEquivK}.

It was originally thought that for orientifolds of the form
$\Omega\times H$, where $H$ is some orbifold group and $\Omega$ worldsheet
parity, 
the D-brane spectrum was classified by Real
equivariant \Ktheory{}~\cite{WittK}, $KR_{\Zop_2 \times
  H}(X)=KO_H(X)$\footnote{We are using the notation where the
  involution appears as part of the group in $KR$-theory. The equality
  between the Real and orthogonal \Kgroup{} follows since the involution
  is taken to have trivial action on the manifold.}. However, it is
sometimes possible to define several distinct orientifolds for a fixed
group $H$; this is somewhat similar to the discrete torsion in
orbifolds. Since the closed string spectra differ for such
orientifolds, the stable D-branes in these backgrounds are also
distinct. It is then clear that $KR_{\Zop_2 \times H}$ cannot describe
the stable D-branes in all such backgrounds.

Our work has been motivated by the $\Omega\times\I_4$ orientifolds.
These were originally studied by~\cite{BianchiSagnotti} before the
discovery of the significance of D-branes~\cite{PolRR}.  It was found
that there were essentially two such models (in the non-compact case);
the massless twisted sector was found to contain either a
hypermultiplet or a tensor multiplet. More recently these orientifolds
were re-considered in D-brane language. In particular the
hypermultiplet model was studied by Gimon and Polchinski~\cite{GP} and
the tensor multiplet model by Blum, Zaffaroni and Dabholkar,
Park~\cite{BZ,DP}.  We will refer to these models throughout the paper
as either hyper, tensor multiplet models or GP, BZDP 
models.\footnote{For a recent review of orientifolds see~\cite{AS}.}

Recently all stable D-branes in the two models have been identified
using BCFT techniques~\cite{QS}. As expected the D-brane spectra are
quite different, and preliminary results suggested that
$KR_{\Zop_2\times\Zop_2}$ corresponds to D-branes in the tensor
multiplet model. It was suggested that D-branes in the hypermultiplet
model should correspond to a twisted version of
$KR_{\Zop_2\times\Zop_2}$ in which the anti-linear $\Omega$ should
{\em anti}-commute with the linear $\I_4$ on the fibres. This proposal
was made by analogy with $\Zop_2\times\Zop_2$ orbifolds. However, it
was unclear whether such an object would in fact form a \Ktheory. In
other words would it satisfy the usual exact sequence and periodicity
properties.

The main goal of this paper is to construct twisted \KRtheories{} for
all consistent orientifolds with fixed group $G$.  In the process we
find a generalisation of group cohomology, so-called group cohomology
with local coefficients, which classifies orientifolds for a given
group $G$. We will find that, just as the different choices in
orbifold theories correspond to projective representations,
orientifolds are classified by projective Real representations. Given
this classification we obtain twisted \KRtheories{} which give the
D-brane spectra of the orientifold theories. In particular we will
apply this construction to find the twisted \KRtheory{} which
classifies stable D-branes in the GP orientifold. This construction
guarantees that the twisted \KRtheories{} satisfy the usual \Ktheory{}
axioms.

In section~\ref{sec:ComputeKOZ2} we compute $KO_{\Zop_2}(\R^{p,q})$
and show that it matches exactly with the D-brane spectrum of the BZDP
model found in~\cite{QS}. Section~\ref{sec:TwistedEquivK} contains
most of our results.  We start by reviewing the construction of
twisted equivariant K--theories in section~\ref{sec:EquivComplexK}. We
generalise this construction to the $\Omega\times\I_4$ orientifolds in
order to find the \KRtheory{} which corresponds to the hypermultiplet
model in section~\ref{sec:DefinitionKR}. We present a general
classification of orientifolds in terms of cohomology with local
coefficients, and the construction of corresponding twisted
\KRtheories{} in section~\ref{sec:OrientifoldClass}.  In
section~\ref{sec:PhysInt} we give a physical interpretation to the
choices allowed for finite abelian orientifolds in terms of phases in
front of the different contributions to one-loop partition functions.
We conclude and present some open problems in
section~\ref{sec:Conclusions}. The paper contains several appendices
where some of the technical details of our calculations are presented.

Some work on the classification of
orientifolds was carried out in~\cite{KleinRab}. In~\cite{SugimotoGB} 
cohomology with local coefficients was discussed 
in a somewhat related, though different, context.

\section{Computation of $KO_{\Zop_2}$}
\label{sec:ComputeKOZ2}

In this section we compute the \Ktheory{} relevant to the non-compact 
BZDP model and show that it agrees exactly with the D-brane spectrum 
found using BCFT techniques. We do this in two different ways; first
we use a long exact sequence similar to the one in~\cite{GS}, then we
show that the result can be easily obtained by using the 
connection between Clifford Algebras and \Ktheory. The former 
method's advantage is that it identifies which D-branes carry the same
charges. This is particularily useful for torsion charged D-branes. The
exact sequence method however, becomes quite cumbersome and it is sometimes
difficult to disentangle the results. 

Following~\cite{GS} we
define a D$p$-brane to be a $(r,s)$-brane if it has $s/r+1$ Neumann 
directions on which $\I_4$ does/does not act and $p=r+s$.
In~\cite{QS} the D-brane spectrum of the BZDP orientifold was computed
using BCFT. We reproduce it here for convenience
\begin{eqnarray}
\label{eq:bzspec}
  \Zop\oplus\Zop 
  &\qquad& \mbox{$r=1,5$ and $s=0,4$}\nonumber \\
  \Zop 
  &\qquad& \mbox{$r=-1,3$ and $s=2$}\nonumber \\
  \Zop 
  &\qquad& \mbox{$r=1,5$ and $s=1,2,3$} \\
  \Zop_2\oplus\Zop_2 
  &\qquad& \mbox{$r=-1,0$ and $s=0$ or $r=3,4$ and $s=4$}\nonumber \\
  \Zop_2
  &\qquad& \mbox{$r=-1,0$ and $s=1$ or $r=3,4$ and $s=3$}\,. \nonumber
\end{eqnarray}
The first two types of D-branes are BPS and are respectively, the
fractional and stuck branes.  The third type of integrally charged
D-brane are the non-BPS truncated branes; the torsion charged branes
are also non-BPS.

In~\cite{QS} it was suggested that $KO_{\Zop_2}$ should be the
\Ktheory{} which classifies such D-branes. In particular, in the
non-compact theory, an $(r,s)$-brane should correspond to
\begin{equation}
  KO_{\Zop_2}(\R^{4-s,5-r})\,,  
\end{equation}
where the bundles are taken to have compact support and $\Zop_2$ acts
as
\begin{equation}
  (x_1, \dots, x_{4-s}, x_{5-s}, \dots, x_{9-s-r}) \mapsto
  (-x_1, \dots, -x_{4-s}, x_{5-s}, \dots, x_{9-s-r})\,.
\end{equation}

As a first step towards computing such KO-theories we note that they 
are 8-periodic
\begin{equation}
  KO_{\Zop_2}(\R^{p,q})=
  KO_{\Zop_2}(\R^{p+8,q})=
  KO_{\Zop_2}(\R^{p,q+8})\,,  
\end{equation}
and that for $p=0$ the group action is trivial and we get immediately
\begin{equation}
  KO_{\Zop_2} (\R^{0,q}) = 
%  KO_\Z{2}^{-q} (\R{0,0}) = 
%  KO_\Z{2}^{-q} ( \{pt\} ) = 
  KO(\R^q) \otimes RO(\Zop_2)\,,  
\end{equation}
where the real representation ring of $\Zop_2$ is
$RO(\Zop_2)=\Zop\oplus\Zop$.  Therefore
\begin{equation}
  \begin{array}{|c||c|c|c|c|c|c|c|c|}
    \hline \strut 
    q & 
    \quad 0 \quad & \quad 1 \quad & \quad 2 \quad & 
    \quad 3 \quad & \quad 4 \quad & \quad 5 \quad &
    \quad 6 \quad & \quad 7 \quad
    \\ \hline \strut
    KO(\R^{q}) & 
    \Zop & \Zop_2 & \Zop_2 & 0 & \Zop & 0 & 0 & 0
%    \\\hline
    \\
    KO_{\Zop_2}(\R^{0,q}) & 
    \Zop\oplus\Zop & \Zop_2\oplus\Zop_2 & \Zop_2\oplus\Zop_2 & 0 & 
    \Zop\oplus\Zop & 0 & 0 & 0
    \\\hline
  \end{array}
\end{equation}
which agrees with the spectrum of $(5-q,4)$-branes.

The other \Kgroups{} can be computed using long exact sequences. For
general manifolds $Y\subset X$ our \Ktheory{} satisfies the usual long
exact sequence 
\begin{equation}
  \dots\rightarrow KO^{-1}_{\Zop_2}(Y)\rightarrow
  KO_{\Zop_2}(X,Y)\rightarrow KO_{\Zop_2}(X)\rightarrow
  KO_{\Zop_2}(Y)\rightarrow \dots
  \,, 
\end{equation}
which, due to periodicity is 24--cyclic. With 
\begin{equation}
  Y=S^{1,0}\times
  \R^{p,q} \subset D^{1,0}\times \R^{p,q}=X\,, 
\end{equation}
this becomes\footnote{$S^{m,n}$ and $D^{m,n}$ are the unit sphere and
  disk in $\R^{m,n}$ with inherited $\Zop_2$--action.}  
\begin{equation}
  \cdots\rightarrow 
  KO^{-1}(\R^{p+q})\rightarrow
  KO_{\Zop_2}(\R^{p+1,q})\rightarrow 
  KO_{\Zop_2}(\R^{p,q})\rightarrow
  KO(\R^{p+q})\rightarrow\cdots
  \label{eq:les1}\,.  
\end{equation}
We have used the fact that $D^{1,0}$ is contractible (in an
$\Zop_2$--equivariant way)
\begin{equation}
  KO_{\Zop_2}(D^{1,0}\times \R^{p,q})=
  KO_{\Zop_2}(\R^{p,q}) 
\end{equation}
and that the $\Zop_2$--action is free on $S^{1,0}=\{+1\}\cup\{-1\}$.
\begin{equation}
  KO_{\Zop_2}(S^{1,0}\times \R^{p,q})=KO(\R^{p+q})\,.  
\end{equation}
Given $KO^\ast_{\Zop_2}(\R^{p,q})$ and $KO^\ast(\R^{p+q})$ it is then
possible to deduce $KO^\ast_{\Zop_2}(\R^{p+1,q})$. Further, for $p=0$ the 
forgetting map 
\begin{equation}
  KO_{\Zop_2}(\R^{0,q})\rightarrow
  KO(\R^{q})
\end{equation}
is onto.  From this it is easy to show that 
\begin{equation}
  \begin{array}{|c||c|c|c|c|c|c|c|c|}
    \hline \strut 
    q & 
    \quad 0 \quad & \quad 1 \quad & \quad 2 \quad & 
    \quad 3 \quad & \quad 4 \quad & \quad 5 \quad &
    \quad 6 \quad & \quad 7 \quad
    \\ \hline \strut
    KO_{\Zop_2}(\R^{1,q}) & 
    \Zop & \Zop_2 & \Zop_2 & 0 & \Zop & 0 & 0 & 0
    \\\hline
  \end{array}
\end{equation}
which is in agreement with the $(5-q,3)$-brane spectrum in
equation~\refeq{eq:bzspec}. Furthermore the maps
\begin{equation}
  KO_{\Zop_2}(\R^{1,s})\rightarrow KO_{\Zop_2}(\R^{0,s}) 
\end{equation}
are one-to-one, indicting that the $\Zop_2$ charge of the
$(r,3)$-brane is the same as (part of the) charge of the $(r,4)$-brane
for $r=3,4$. One may continue in this way, however it becomes
increasingly difficult to solve the extension ambiguities.

Instead we will now use the connection between \Ktheory{} and Clifford
algebras to compute the \Kgroups. One finds that~\cite{VBthesis}
\begin{equation}
  KO_{\Zop_2}(\R^{p,q}) = KO^{(0,p+1)}(\R^q)\,.
\end{equation}
In Appendix~\ref{appendix:Clifford} we define $KO^{(m,n)}(X)$ and
prove the above result.  The above formula is useful since the object
on the right hand side is purely algebraic and is well 
known~\cite{KaroubiBook}:
\begin{equation}
  KO_{\Zop_2}(\R^{p,q}) = 
  \vcenter{
  \xymatrix@=0.2cm{
   {\scriptstyle q=7} & 
   0 & 0 & 0 & 0 & 0 & 0 & 0 & 0 \\
   {\scriptstyle q=6} & 
   0 & 0 & \Zop & \Zop_2 & \Zop_2\oplus\Zop_2 & \Zop_2 & \Zop & 0 \\
   {\scriptstyle q=5} & 
   0 & 0 & 0 & \Zop_2 & \Zop_2\oplus\Zop_2 & \Zop_2 & 0 & 0 \\
   {\scriptstyle q=4} & 
   \Zop\oplus\Zop & \Zop & \Zop & \Zop & 
   \Zop\oplus\Zop & \Zop & \Zop & \Zop \\
   {\scriptstyle q=3} & 
   0 & 0 & 0 & 0 & 0 & 0 & 0 & 0 \\
   {\scriptstyle q=2} & 
   \Zop_2\oplus\Zop_2 & \Zop_2 & \Zop & 0 & 0 & 0 & \Zop & \Zop_2 \\
   {\scriptstyle q=1} & 
   \Zop_2\oplus\Zop_2 & \Zop_2 & 0 & 0 & 0 & 0 & 0 & \Zop_2 \\
   {\scriptstyle q=0} & 
   \ar[]+/d 0.5cm/+/l 1.0cm/;[uuuuuuu]+/d 0.5cm/+/l 1.0cm/+/u 0.8cm/  
   \ar[]+/d 0.5cm/+/l 1.0cm/;[rrrrrrr]+/d 0.5cm/+/l 0.6cm/+/r 1cm/
   \Zop\oplus\Zop & \Zop & \Zop & \Zop & 
   \Zop\oplus\Zop & \Zop & \Zop & \Zop \\
   & {\scriptstyle p=0} 
   & {\scriptstyle p=1} 
   & {\scriptstyle p=2} 
   & {\scriptstyle p=3} 
   & {\scriptstyle p=4} 
   & {\scriptstyle p=5} 
   & {\scriptstyle p=6} 
   & {\scriptstyle p=7} 
  }}
\end{equation}
Comparing with equation~\refeq{eq:bzspec} we see exact agreement between the
spectrum of BZDP D-branes and the \Ktheory{} predictions.

\section{Twisting in equivariant \Ktheory}
\label{sec:TwistedEquivK}

In this section we construct a \Ktheory{} which describes the D-brane
spectrum of the GP orientifold. Comparing the D-brane spectrum of the
GP model (see appendix~\ref{appendix:GPorientifold}) with the results
of the previous section it is clear that this is not described by
$KO_{\Zop_2}$.  Instead, we argue that D-branes in the hyper model are
described by a twisted \Ktheory{} $KR_{\Zop_2\times\Zop_2}^{[D_8]}$.
We present a unified picture of twisting equivariant \Ktheories, which
allows for twists involving linear as well as anti-linear group
elements. We begin the section by discussing such twisting in the case
of $K_{\Zop_2\times\Zop_2}^{[D_8]}$ which gives the D-brane spectrum
of the $\Zop_2\times\Zop_2$ orbifold with discrete torsion. In the
following subsection we obtain $KR_{\Zop_2\times\Zop_2}^{[D_8]}$. The
construction is then generalised to describe generic non-compact
orientifolds; as a by-product we show that these are classified by a
cohomology group $H^2\big(*,\Ut\big)$ much in the same way as
orbifolds are classified by $H^2(*,\Uone)$~\cite{VafaT}.

\subsection{Twisting for ordinary orbifolds}
\label{sec:twistfordummies}

Let us quickly review how group cohomology enters twists of ordinary
\Ktheory{}, that is how discrete torsion alters the \Kgroups{}. 
This section is not strictly necessary for the understanding of the
rest of the paper, we only want to recall how the twisted \Ktheory{}
appearing in the analysis of WZW models
\cite{SchomerusWZWa,SchomerusWZWb} is related to discrete torsion
and projective representations.

If spacetime $X$ is a bona fide manifold, i.e. not an orbifold, then
the twist is caused by a nontrivial B-field. In such a background 
\Dbranes{} no longer carry ordinary gauge bundles but ``twisted
bundles'', where the transition functions $h_{ij}$ do not close
\cite{WittK,FreedWitten,BouwknegtMathai}.  Instead there is a $\Uone$
valued function on triple overlaps such that
\begin{equation}
  h_{jk} h^{-1}_{ik} h_{ij} = g_{ijk}\,.
\end{equation}
Ignoring torsion in $H^3(X,\Zop)$ the twist class corresponds to the
de Rahm class of the flux $\diff B$ via the well-known
identity\footnote{$\underline{\Uone}$ is the sheaf of $\Uone$ valued
  functions.}
\begin{equation}
  [g_{ijk}]\in 
  H^2\left(X;\underline{\Uone}\right)
%  \stackrel{\displaystyle \sim}{\longrightarrow}
  \overset{\sim}{\longrightarrow}
  H^3(X;\Zop)
  \owns
  [\diff B]\,.
\end{equation}
Mathematically this corresponds to the statement that you can twist
ordinary \Ktheory{} $K(X)$ by $H^2\big(X;\underline{\Uone}\big)$.

Now orbifold string theory on $X/G$ is really $G$-equivariant theory
on the manifold $X$. The relevant \Ktheory{} is equivariant \Ktheory{}
$K_G(X)$ and it can be twisted by classes in the equivariant
cohomology group $H^2_G\big(X;\underline{\Uone}\big)$. Now for general
(topologically nontrivial) manifolds $X$ all possible twists may be
difficult to determine, but there is a rather well-understood subclass
of twists. These come from pullbacks via the projection $\pi:X\to
\mathrm{pt}$ from $H^2_G(\mathrm{pt};\Uone)$, that is we are
interested in the image
\begin{equation}
  \pi^\ast\Big( H^2_G(\mathrm{pt};\Uone) \Big) 
  \subset 
  H^2_G\big(X;\underline{\Uone}\big)\,.
\end{equation}
The advantage of this subclass is that
\begin{equation}
  H^2_G(\mathrm{pt};\Uone) = H^2(BG; \Uone) = H^2(G,\Uone)\,,
\end{equation}
where $BG$ is the base space of the classifying space $EG$.
The connection to group cohomology yields an interpretation
of spacetime twists as projective representations. Motivated
by this we now turn to the classification of projective representations.

\subsection{Twisted equivariant complex \Ktheory}
\label{sec:EquivComplexK}

For an orbifold group $G$ which admits projective representations (see
below) there are several orbifolds consistent with modular
invariance~\cite{VafaT}. Typically it is possible to define the action
of some generator $g_i$ in several different ways on a $g_j$-twisted
sector ($i\neq j$) giving distinct closed string backgrounds.  As a
result the D-brane spectrum is distinct in each of the orbifolds.
$G$-equivariant \Ktheory{} will then describe the D-brane spectrum in
the orbifold without discrete torsion, and one has to define
twisted versions of $K_G$ which describe the D-brane spectrum of
the various orbifolds with torsion. These twisted \Ktheories{} are the
Grothendieck group of isomorphism classes of bundles with a projective
representation of $G$ on the fibres rather than a proper
representation. 

Recall that a projective representation of a finite group $G$ is a
representation of the central extension of $G$ by $\Uone$ such that 
$\Uone$ acts by multiplication with a phase. In other words a
projective representation of $G$ is a choice of $H$ such that the
following sequence is exact 
\begin{equation}
  \label{eq:U1extension}
  1
  \rightarrow
  \Uone
  \stackrel{i}{\longrightarrow}
  H
  \stackrel{\pi}{\longrightarrow}
  G
  \rightarrow 
  1  
\end{equation}
and $\Uone$ is in the centre of $H$. 

Choose a section $s:G\to H$ such that $\pi \circ s = \Id{G}$. This is
always possible since $\pi$ is surjective, but in general $s$ will not
be a group homomorphism. The failure to be a group homomorphism
\begin{equation}
  s(g_1) s(g_2) = c(g_1,g_2) s(g_1 g_2)  
\end{equation}
defines a function $c:G\times G\to \Uone$. An ordinary representation of $H$
$\rho:H\to \GL_n$ defines a projective representation of $G$ via
$\gamma = \rho \circ s: G\to \GL_n$, which will be a ``representation
up to phases''. In particular, as is familiar to physicists, it
satisfies
\begin{equation}
  \gamma(g_1)\gamma(g_2)=c(g_1,g_2)\gamma(g_1g_2)\,.
\end{equation}
Requiring that $s$ or $\gamma$ be associative restricts $c$ to 
satisfy
\begin{equation}
  \label{eq:coboundary2}
  c(g_1, g_2) 
  \frac{1}{c(g_1, g_2 g_3)}
  c(g_1 g_2 , g_3) 
  \frac{1}{c(g_2,g_3)} = 1\,.
\end{equation}
In group cohomology language the left hand side defines the coboundary
of a two-cocycle and the above equality says that $c$ is coclosed.
Further, given a function $G\to \Uone$ (by abuse of notation also
called $c$) we can replace $s(g)\to c(g) s(g)$ and then
\begin{equation}
  s(g_1) s(g_2) = c(g_1,g_2) s(g_1 g_2)  
  \quad \rightarrow \quad
  s(g_1) s(g_2) = 
  \frac{c(g_1 g_2)}{c(g_1) c(g_2)}
  c(g_1, g_2)  s(g_1 g_2)\,.
\end{equation}
Hence $2$-cocycles that differ by 
\begin{equation}
  \label{eq:coboundary1}
  c(g_1) c(g_1 g_2)^{-1} c(g_2)
\end{equation}
correspond to the same extension. The above defines a coboundary of a
$1$-cochain in group cohomology, and therefore we identify
\begin{equation}
  \label{eq:centralextcohomology}
\left\{ 
    \begin{array}{c}
      \text{projective} \\
      \text{representations of $G$}   
    \end{array}
  \right\}
 \quad = \quad
  \left\{ 
    \begin{array}{c}
      \text{central extensions} \\
      1  \rightarrow  \Uone
      \rightarrow  H \rightarrow  G  
      \rightarrow  1        
    \end{array}
  \right\}
  \quad = \quad
  H^2\big(G,\Uone\big)\,.
\end{equation}
A twisted $G$-equivariant vector bundle is a vector bundle with the
group $G$ acting by a projective representation.\footnote{One of us
  (BS) learnt about this approach to twisted equivariant \Ktheory{}
  from Burt Totaro in July 2000~\cite{Totaropriv}. We have later been informed
  that it has been known to others~\cite{Atiyahpriv,Segalpriv}.}  Such bundles
form a semigroup under the Whitney sum, and as usual we can define the
Grothendieck group $K_G^{[H]}(X)$ corresponding to $H\in
H^2\big(G,\Uone\big)$. By equation~(\refeq{eq:centralextcohomology}) 
any representation of $H$ is either a proper or a projective
representation of $G$. Succinctly, at the level of K-theory this implies
a decomposition
\begin{equation}
\label{eq:equivCtw}
K_H(X)=
K_G(X)\oplus K^{[H]}_G(X)\,.
\end{equation}
Since both $K_G$ and $K_H$ satisfy the usual \Ktheory{} properties, such
as periodicity and long exact sequences, then so does $K^{[H]}_G$.

As an example consider $G=\Zop_2\times \Zop_2$ with generators 
$g_1,g_2$. The D-brane spectrum of these orbifolds is well
known~\cite{SCY,GabCY}. The groups $K_{\Zop_2\times\Zop_2}$ for
Euclidean space have been computed and have been found to
agree with the D-brane spectrum of the orbifold with no discrete
torsion~\cite{SCY}. Since
\begin{equation}
  \label{eq:groupcohomologyusual}
  H^2\big(\Zop_2\times \Zop_2,\Uone\big)=\Zop_2
\end{equation}
there is one nontrivial projective representation given by the
following choice of normalised (meaning $s(1)=1$) section:
\begin{equation}
  s(g_1) s(g_2) = - s(g_1 g_2) = - s(g_2) s(g_1)
  , \quad
  s(g_1)^2 = s(g_2)^2 = 1\,.
\end{equation}
This projective representation is a representation of the group $D_8$
\begin{equation}
  D_8 = \Big\{ 
  \sigma, \gamma_1, \gamma_2 
  \,\Big| \,
  \gamma_1 \gamma_2 = \sigma \gamma_2 \gamma_1,~
  \gamma_1^2=\gamma_2^2=\sigma^2=1
  \Big\}
\end{equation}
where the generator $\sigma$ is represented by $-1$ (and acts
trivially on the base space $X$). Since $D_8$ irreducible representations 
decompose into projective and ordinary representations of
$\Zop_2\times \Zop_2$ the \Ktheory{} of $D_8$ splits as in 
equation~\refeq{eq:equivCtw}:
\begin{equation}
  \label{eq:KD8decompose}  
  K_{D_8}(X)=
  K_{\Zop_2\times\Zop_2}(X) \oplus 
  K^{[D_8]}_{\Zop_2\times\Zop_2}(X)\,.
\end{equation}
The above equation not only shows that $K^{[D_8]}_{\Zop_2\times\Zop_2}$
is well defined but is also useful in computing twisted equivariant
\Kgroups{} if the groups act trivially on the base (e.g.
$X=\mathrm{pt}$).  As usual we have
\begin{equation}
  \label{eq:KpointRepresentation}
  K_G^i(\text{pt}) = K^i(\text{pt}) \otimes R[G]\,,
\end{equation}
where the representation rings are~\cite{Serre} 
\begin{equation}
  R[D_8]=\Zop^{\oplus 5}
  \,,\qquad 
  R[\Zop_2\times\Zop_2]=\Zop^{\oplus 4}\,.
\end{equation}
Then equation~(\ref{eq:KD8decompose}) yields\footnote{Below, and 
throughtout the text, we write equalities between different K-groups. These
are meant to indicate that there is an isomorphism between such groups. It
should be however, understood that such equalities need not map generators to
generators; for example presently in the equality between
$K^{[D_8],i}_{\Zop_2\times\Zop_2}(\text{pt})$ and $ K^i(\text{pt})$ the
generator of the twisted group gets mapped to twice the generator of the
complex K--theory.}
\begin{multline}
  K^i(\text{pt}) \otimes \Zop^5 = 
  \Big( K^i(\text{pt}) \otimes \Zop^4 \Big) \oplus
  K^{[D_8],i}_{\Zop_2\times\Zop_2}(\text{pt})
  \\ \Rightarrow \quad
  K^{[D_8],i}_{\Zop_2\times\Zop_2}(\text{pt}) = K^i(\text{pt})
  = \left\{  
    \begin{array}{cl}
      0 & i~\text{odd} \\
      \Zop & i~\text{even}
    \end{array}
  \right.
\end{multline}
in agreement with the presence of $(2k_1+1;2k_2,2k_3,2k_4)$-branes in
the Type IIB orbifold with discrete torsion~\cite{GabCY}. 

We note that the twisted \Ktheory{} defined by representations of $Q_8$,
the unit quaternions, 
\begin{equation}
  Q_8 = \Big\{ 
  \sigma, \gamma_1, \gamma_2
  \,\Big| \,
  \gamma_1 \gamma_2 = \sigma \gamma_2 \gamma_1,~
  \gamma_1^2=\gamma_2^2=\sigma, \sigma^2=1
  \Big\}
\end{equation}
is the same as the $D_8$-twisted \Ktheory{} defined above. This is
because the two yield projective representations of
$\Zop_2\times\Zop_2$ that differ by a coboundary (\cf
equation~\refeq{eq:groupcohomologyusual}). This will be in contrast to the
case of Real \Ktheory, as we will see in the next section.

\subsection{Defining $KR_{\Zop_2\times\Zop_2}^{[D_8]}$}
\label{sec:DefinitionKR}

In this subsection we wish to extend the construction of twisted
equivariant \Kgroups{} to \KRgroups. Following the above discussion it
seems clear that D-branes in the GP model are described by a
\Ktheory{} of bundles on which $g$ acts linearly, $\tau$ acts
anti-linearly, and the two anti-commute. Consider then $D_8$-equivariant
KR theory in which $g$ and $\sigma$ act linearly and $\tau$ acts
anti-linearly. Such \Ktheories{} have been considered in the
mathematical literature~\cite{KaroubiSeminar,AtiyahSegalEquiv}. Any
representation\footnote{We use the word representation somewhat
  loosely here as $\gamma(\tau)$ acts anti-linearly on a
  vector space. Really we are talking about Real representations; we
  shall give more precise definitions in the next section.} of $D_8$
is either a representation of $\Zop_2\times\Zop_2$ or a projective
representation ($\tau$ and $g$ anti-commute). At the level of
\Ktheory{} this is
\begin{equation}
  \label{eq:KRD8decompose}  
  KR_{D_8}(X)=
  KR_{\Zop_2\times\Zop_2}(X) \oplus 
  KR^{[D_8]}_{\Zop_2\times\Zop_2}(X)\,.
\end{equation}
As in the previous section this guarantees that
$KR^{[D_8]}_{\Zop_2\times\Zop_2}$ is a \Ktheory.

In order to confirm that $KR^{[D_8]}_{\Zop_2\times\Zop_2}$ is the
\Ktheory{} which describes D-branes in the hypermultiplet 
model we shall compute
it on $\R^{0,i}$, on which $\Zop_2\times\Zop_2$ acts trivially. There
are three irreducible representations of $D_8$ with $\tau$ acting
anti-linearly (we denote complex conjugation by $\delta$)
\begin{equation}
  \begin{array}{|c||c|c|c|c|}
    \hline \strut 
    g & 
    \quad 1 \quad & \quad \sigma \quad & 
    \quad g \quad & \quad \tau \quad 
    \\ \hline \strut
    r_1(g) & 
    \begin{pmatrix} 1 \end{pmatrix} &
    \begin{pmatrix} 1 \end{pmatrix} &
    \begin{pmatrix} 1 \end{pmatrix} &
    \begin{pmatrix} 1 \end{pmatrix}\circ \delta 
    \\\hline
    r_2(g) & 
    \begin{pmatrix} 1 \end{pmatrix} &
    \begin{pmatrix} 1 \end{pmatrix} &
    \begin{pmatrix} -1 \end{pmatrix} &
    \begin{pmatrix} 1 \end{pmatrix}\circ \delta
    \\\hline
    r_3(g) & 
    \begin{pmatrix}  1 & 0 \\ 0 &  1 \end{pmatrix} &
    \begin{pmatrix} -1 & 0 \\ 0 & -1 \end{pmatrix} &
    \begin{pmatrix}  0 & i \\ -i &  0 \end{pmatrix} &
    \begin{pmatrix}  1 & 0 \\ 0 &  1 \end{pmatrix} \circ \delta
    \\\hline
  \end{array}
\end{equation}
The analogue of equation~\refeq{eq:KpointRepresentation} in \KRtheory{}
is~\cite{AtiyahSegalEquiv}
\begin{equation}
  \label{eq:KRGdecompose}
  KR_G(X) = 
  \Big( A_G\otimes KO(X) \Big) \oplus
  \Big( B_G\otimes K(X) \Big) \oplus
  \Big( C_G\otimes KSp(X) \Big)\,, 
\end{equation}
where the representation ring of $G$ decomposes as
\begin{equation}
  R[G] = A_G \oplus B_G \oplus C_G
\end{equation}
according to commuting field $\R,\C,\Hbb{}$.  Both $1$-dimensional
representations have commuting field $\R$ and one can easily show that
the commuting field for $r_3$ is $\C$.\footnote{The (complex) matrices
  commuting with $r_3$ are
  \begin{equation}
    \mathbb{F}_3 \eqdef 
    \R
    \begin{pmatrix} 1 & 0 \\  0 & 1 \end{pmatrix} +
    \R
    \begin{pmatrix} 0 & 1 \\  -1 & 0 \end{pmatrix}
    \simeq \C
  \end{equation}
  so the commuting field is $\C$.}  Then equations~\refeq{eq:KRD8decompose}
and \refeq{eq:KRGdecompose} gives
\begin{equation}
  \left.
  \begin{array}{l}
    KR_{\Zop_2\times\Zop_2}(X) = KO(X)\oplus KO(X) \\
    KR_{D_8}(X) = KO(X)\oplus KO(X) \oplus K(X)
  \end{array}
  \right\} 
  \Rightarrow~
  KR_{\Zop_2\times\Zop_2}^{[D_8]}(X)=K(X)
\end{equation}
In particular for $X=\R^i$ with trivial group action we find that
\begin{equation}
  KR_{\Zop_2\times\Zop_2}^{[D_8],i}(\text{pt})=K^i(\text{pt})
  = \left\{  
    \begin{array}{cl}
      0 & i~\text{odd} \\
      \Zop & i~\text{even}
    \end{array}
  \right.
\end{equation}
which is in agreement with the presence of $\Zop$ charged
(2k+1,4)-branes in the GP model~\cite{QS} (see also~\refeq{eq:Dbranespec}).

Repeating the above construction for the unit quaternions $Q_8$ to obtain
$KR_{\Zop_2\times\Zop_2}^{[Q_8]}$ one comes across a surprise. As we show 
in appendix~\ref{appendix:GPorientifold}
\begin{equation}
KR_{\Zop_2\times\Zop_2}^{[Q_8]}(\mathrm{pt})=
KSp_{\Zop_2}(\mathrm{pt})\neq 
KR_{\Zop_2\times\Zop_2}^{[D_8]}(\mathrm{pt})\,.
\end{equation}
This stems from the fact that projective representations of 
$\Zop_2\times\Zop_2$ which are representations of $D_8$
(with $\tau$ acting anti-linearly) are not equivalent modulo coboundaries
to those of $Q_8$. Clearly then we need a generalisation of $H^2(G,\Uone)$
to classify all inequivalent such representations.

\subsection{Classification of Orientifolds}
\label{sec:OrientifoldClass}

In the previous subsection we have shown that projective
representations of $\Zop_2\times\Zop_2$ in which one of the generators
acts anti-linearly differ significantly from linear projective
representations. In this subsection we will generalise group
cohomology to take into account such differences. This will allow us
to classify the analogue of discrete torsion in orientifolds, and to
obtain suitable twistings of \KRtheory{} which will describe D-branes in
such models.

An orientifold group has linear and anti-linear elements.
To keep track of which act linearly and which anti-linearly
we will use the notion of an {\em augmented} group, that is a group
together with a homomorphism $\epsilon:G\to \Zop_2$. A Real
representation of $G$ on some complex vector space $V$ associates
to each $g\in G$ a linear or antilinear map $V\to V$ depending on
whether $\epsilon(g)=+1$ or $\epsilon(g)=-1$. 

As before we want $G$ to act ``up to phases''. By the same reasoning
as in section~\ref{sec:EquivComplexK} this means we have to find an
extension
\begin{equation}
  \label{eq:U1extensionReal}
  \xymatrix{
  1 \ar[r] & 
  \Uone \ar[r]^i & 
  H \ar[r]^\pi \ar[dr]_{\epsilon'\eqdef \epsilon \circ \pi} & 
  G \ar[r] \ar[d]^\epsilon & 
  1 
  \\   &    &    &   \Zop_2 &  \\
  }
\end{equation}
However, there are two important differences compared to
equation~\refeq{eq:U1extension}:
\begin{itemize}
\item $G$ is now an augmented group, and $H$ inherits an augmentation
  $\epsilon'$.
\item The extension is no longer central: complex conjugation acts on
  the $\Uone$. 
\end{itemize}
Since anti-linear elements act by complex conjugation on
$\Uone$-phases, conditions~\refeq{eq:coboundary2}
and~\refeq{eq:coboundary1} have to be modified. In particular the
differentials of $1$- and $2$-cochains are
\begin{subequations}
\begin{align}
  \label{eq:coboundaryReal}
  &\big(\diff c\big) (g_1,g_2) =
  c(g_1) \frac{1}{c(g_1 g_2)} g_1\circ c(g_2)
  \\ 
\label{eq:coboundaryReal3}
  &\big(\diff c\big) (g_1,g_2,g_3) =
  c(g_1, g_2) 
  \frac{1}{c(g_1, g_2 g_3)}
  c(g_1 g_2 , g_3) 
  \frac{1}{g_1\circ c(g_2,g_3)}
\end{align}
\end{subequations}
where 
\begin{equation}
  g\circ c(h) \eqdef
  \begin{cases}
    \overline{c(h)} & \text{if}~\epsilon(g)=1 \\
    c(h) & \text{if}~\epsilon(g)=0 
  \end{cases}
\end{equation}
Mathematically this is well-known as group cohomology with local
coefficients (by abuse of notation again denoted $H^\ast(G,F)$) where
the group in the first slot, $G$, acts on the second, $F$.\footnote{
In~\cite{SharpeDT} it was noted that this possibility so far did not
appear in the physics literature, but the case at hand shows that it
is indeed necessary.} We will use $\Ut$ to denote the ``$\Uone$ with
action on it''. Then
\begin{equation}
  H^2\left(G, \Ut\right)
\end{equation}
classifies all inequivalent non-compact $G$ orientifolds. Further a
non-trivial projective Real representation of $G$ gives a Real
representation of some group $H$ and hence an element $[H]\in
H^2\big(G, \Ut\big)$, just as in equation~\refeq{eq:centralextcohomology}.
This may be used to construct $ KR_G^{[H]}$, the \Ktheory{} which
gives the D-brane spectrum of this particular $G$ orientifold
\begin{equation}
  KR_H(X)=KR_G(X)\oplus KR_G^{[H]}(X)\,.
\end{equation}

In appendix~\ref{appendix:generalgroup} we compute the cohomology of the
most general finite abelian orientifold group, in particular we find
\begin{equation}
 H^2\left(\Zop_2\times\Zop_2, \Ut\right)=\Zop_2\oplus\Zop_2\,.
\end{equation}
As we have seen the projective Real representations given by $[Q_8]$
and $[D_8]$ are inequivalent, and so they can be taken as the
generators of $H^2\big(\Zop_2\times\Zop_2, \Ut\big)$.  From the
explicit $2$-cocycles (see
appendix~\ref{appendix:explicitlocalcohomology}) we can then identify
the following inequivalent projective Real representations of
$\Zop_2\times\Zop_2$,
\begin{equation}
\begin{aligned}{} %dont delete the empty {} 
  [(\Zop_2)^3] &:\quad g^2 =1,~ \tau^2 = 1,~ g\tau=\tau g\,, \\
  [D_8]        &:\quad g^2=1,~ \tau^2=1,~ g\tau=-\tau g\,, \\
  [Q_8]        &:\quad g^2=-1,~ \tau^2=-1,~ g\tau=-\tau g\,, \\
  [\Zop_2\times\Zop_4]    &:\quad g^2=-1,~ \tau^2=-1,~ g\tau=\tau g\,. \\
\end{aligned}
\end{equation}
As a result there are four inequivalent twisted
$KR_{\Zop_2\times\Zop_2}$ theories. Those
twisted by $(\Zop_2)^3[]$ (i.e. untwisted) and $[D_8]$ give the
D-brane spectrum of the tensor and hyper models. Space-time filling branes
in the $[Q_8]$ twisted theory are classified by $KSp_{\Zop_2}$. This
describes D-branes in the $\I_4$ orbifold of the Type I theory with
$Sp$ gauge group, and a twsited sector tensor multiplet. 
Similarly then the theory
twisted by $[\Zop_2\times\Zop_4]$ gives the D-brane spectrum of the $\I_4$
orbifold of the Type I theory with $Sp$ gauge group, and a twisted sector
hypermultiplet.  In appendix~\ref{appendix:GPorientifold}, 
generalising~\cite{QS}, we have computed the D-brane
spectrum of both the $Sp$ orientifolds, and (partially) matched it
with the corresponding twisted \KRtheories.

As a further example to illustrate the construction 
consider orientifolding Type IIB by $\Omega$. One easily shows that
\begin{equation}
  H^2\left(\Zop_2, \Ut\right)=\Zop_2\,.
\end{equation}
The untwisted \KRtheory{} is simply
\begin{equation}
  KR_{\Zop_2}(X)=KO(X)\,,
\end{equation}
which indeed describes D-branes in the Type I $SO$ theory. The
\KRtheory{} twisted by the generator of $H^2\big(\Zop_2, \Ut\big)$
is
\begin{equation}
  KR_{\Zop_2}^{[\Zop_4]}(X)=KSp(X)\,,
\end{equation}
which gives the D-brane spectrum of the Type I $Sp$ theory. This example
was also discussed in~\cite{Gukov}.

\section{Physical interpretation}
\label{sec:PhysInt}

In the previous section we have shown that for an orientifold group
$G$ there are $H^2(G,\Ut)$ different models, and that given a
particular such orientifold $[H]\in H^2(G,\Ut)$ we can construct the
\Ktheory{} $KR^{[H]}_G$ which classifies the stable D-branes in it.
In appendix~\ref{appendix:generalgroup} we have computed
$H^2(G,\Ut)$ for the most general finite abelian orientifold group. In
this section we analyse the various one-loop partition functions in
the orientifold and identify the places where a choice of phase is
allowed without spoiling the properties of such partition functions.
We will show that the physically acceptable choices are isomorphic to
elements of $H^2(G,\Ut)$.

The most general finite abelian orientifold group $G$ is generated by
anti-linear elements $t_1,\cdots,t_a$ and linear elements
$s_1,\cdots,s_b$. This is equivalent to the orientifold group
generated by $t,g_1,\cdots,g_n,h_1,\cdots,h_m$, where $t$ is the only
anti-linear element (of even order), the $g_i$ are linear even-order
elements and the $h_i$ are linear odd-order elements.\footnote{This
  equivalence holds as $<t_1,\cdots,t_a>$ generates the same group as
  $<t_1,t_1t_2\cdots,t_1t_a>$ and $t_1t_i$ is a linear even-order
  element.} In appendix~\ref{appendix:generalgroup} we show that
\begin{equation}
  \label{eq:genabcohom}
  H^2(G,\Ut)=
  \Big( \mathop{\oplus}_{i=1}^{n} \Zop_2 \Big) \oplus
  \Zop_2 \oplus
  \Big( \mathop{\oplus}_{i<j=1}^n \Zop_2 \Big) =
  \Zop_2^{\oplus n(n+1)/2+1}\,.
\end{equation}
We will interpret each of the three terms in the middle of the above
equation as coming from phase choices in front of the various one-loop
partition functions.

Consider first the torus amplitude. In an orbifold
given two generators $K$ and $L$ of order $k$ and $l$ it is possible
to choose a phase $\omega^p=\exp(2\pi i p/\gcd(k,l))$ with
$p=1,\cdots,\gcd(k,l)$ in front of the torus amplitude
\begin{equation}
  \omega^p\Tr^{K}\left(Le^{-tH_c}\right)\,.
\end{equation}
The trace is taken over the $K$-twisted sector with $L$ inserted and
$H_c$ is the closed string Hamiltonian. This phase effectively changes
the action of $L$ on the $K$-twisted groundstate. Then for $K$ and $L$
to remain order $k$ and $l$ respectively, various other parts of the
torus amplitude will change their phases. For example we will have
\begin{equation}
  \omega^{-p}\Tr^{K}\left(Le^{-tH_c}\right)\,.
\end{equation}
Recently it has been shown~\cite{KleinRab} that orbifolds with
discrete torsion different from $\pm 1$ cannot be consistently
projected by $\Omega$. The argument also applies to more general
anti-linear elements $t$. From it we see that the $\oplus_{i<j=1}^n
\Zop_2$ factor in equation~\refeq{eq:genabcohom} comes from the
orbifold discrete torsion which is allowed for an orientifold
background.  Hence, as in~\cite{VafaT}, the phase in front of the
$g_1$-twisted sector amplitude with $g_2$ inserted is proportional to
$c(g_1,g_2)/c(g_2,g_1)$ where $c\in H^2\big(G,\Ut\big)$. In particular
it is worth noting that there is no consistent discrete torsion
between two odd-order elements.

An anti-linear order two element $\tau\in G$ gives rise to an
Orientifold plane coupling to the untwisted sectors.\footnote{For
  example $\Omega\I_n$ gives rise to an O$(9-n)$-plane described by
  the crosscap state
  \begin{equation}
    \ket{O(9-n)}=\ket{C(9-n)}_{\mbox{\scriptsize\NSNS}}+
    \ket{C(9-n)}_{\mbox{\scriptsize\RR}}\,.
  \end{equation}
} As is well known the overall sign of the normalisation of this
crosscap state can be freely chosen; for example this choice of sign
in the $\Omega$ orientifold of Type IIB gives the Type I theory
with $SO$ or $Sp$ gauge groups. It is easy to show that for
anti-linear $\tau\in G$ with $\tau^2=1$ $c(\tau,\tau)=\pm
1$\footnote{With  $g_1=g_2=g_3=\tau$ the cocycle 
condition~\refeq{eq:coboundaryReal3} becomes
\begin{equation}
c(\tau,\tau)c(1,\tau)=c(\tau,1){\bar c(\tau,\tau)}\,.
\end{equation}
Since $c(1,\tau)=c(\tau,1)=1$ the above implies $c(\tau,\tau)=
{\bar c(\tau,\tau)}=\pm 1$.} and as a result the corresponding M\"obius
strip amplitude has the phase
\begin{equation}
  c(\tau,\tau)\Tr\left(\tau e^{-tH_o}\right)\,,
\end{equation}
where $H_o$ is the open string Hamiltonian. Such phase choice is
possible for $t\in G$ as well as for $tg_i\in G$ if the order of these
elements is $4k+2$.

For an anti-linear element $\tau\in G$ of order $4k$ the difference of
signs between the Klein bottle amplitudes
\begin{equation}
  \Tr^{\tau^2}\left(\tau e^{-tH_c}\right)
  \qquad\text{and}\qquad 
  \Tr^{\tau^2}\left(\tau^{2k+1} e^{-tH_c}\right)
\end{equation}
gives two different, consistent models (see for
example~\cite{RabUranga})\footnote{We thank A. Uranga for a discussion on 
this.}.  With a bit more work it is possible to show as above that
$c(\tau,\tau)/c(\tau^{2k+1},\tau^{2k+1})=\pm 1$, and this is the
cocycle contribution which keeps track of this choice.

We can now explain the appearance of
$\big(\negmedspace\oplus_{i=1}^{n}\Zop_2\big)\oplus\Zop_2$ in
equation~\refeq{eq:genabcohom}. Each element $t,tg_1,\cdots,tg_n$ is
even-order and anti-linear. If its order is $4k+2$ then we may choose
a phase proportional to $c(tg_i,tg_i)$ which governs the overall sign
of the orientifold plane. On the other hand if it is of order $4k$ we
may choose a sign proportional to
$c(tg_i,tg_i)/c((tg_i)^{2k+1},(tg_i)^{2k+1})$ as described in the
previous paragraph. Either way each $t,tg_1,\cdots,tg_n$ gives rise to
a $\Zop_2$ choice. In total this reproduces the
$\big(\negmedspace\oplus_{i=1}^{n}\Zop_2\big)\oplus\Zop_2$ factor in
equation~\refeq{eq:genabcohom}.

Finally, it is possible to convince oneself that there are no other
phase choices that we can make consistently. For example in the
$\Omega\times\Zop_3$ orientifold we may only choose the overall sign of
the O9-planes. One might think that the 
natural phase $e^{2\pi i/3}$ could appear in
the twisted sector crosscaps. However, the square of this phase appears
in the untwisted sector Klein bottle with $g$ (the generator of $\Zop_3$)
or $g^2$ inserted.  The action of $g$ on the untwisted sector is
unique and hence we cannot pick up a phase here. This argument can be
extended to show that indeed the above choices are the only ones which
are consistent.

\section{Conclusion and Outlook}
\label{sec:Conclusions}

We have computed $KR_{\Zop_2\times\Zop_2}(\R^{p,q})$ which classifies
D-branes in the $\I_4\times\Omega$ orientifold with twisted sector
tensor multiplet, and found complete agreement with the BCFT
results~\cite{QS}. We have also constructed a twisted version of this
KR--theory relevant to the model with twisted sector hypermultiplet. In
the process we have identified a type of cohomology which classifies
orientifolds, in a similar way to the classification of orbifolds by
the second group cohomology~\cite{VafaT}.  We have presented a unified
approach towards twisting complex and Real \Ktheories. This procedure
allows for the straightforward identification of \Kgroups{} relevant
to orbifolds and orientifolds with discrete torsion. We have found
places in the various one-loop diagrams where $\pm 1$ phases may be
introduced and have shown that for finite abelian orientifold groups
this freedom is precisely described by cohomology with local
coefficients.

In compact orientifolds it was shown that not all possible
orientifolds are allowed~\cite{PolT}. In particular there are global
conditions which only allow configurations with $8k$ hypermultiplets
and $32-8k$ tensor multiplets $(k=0,1,2,3,4)$. It would be very
interesting to show that these results follow from the cohomology we
have presented here. Perturbative orientifolds with the same
orientifold group differ from one another by the presence of discrete
background $B_{\mu\nu}$ fields, it should be possible to make this
connection more precisely. In particular it would be interesting to
understand better how the ten-dimensional Type I $SO$ and $Sp$
theories are connected.  Finally, it should prove instructive to try
to obtain the classification by cohomology with local coefficients
from considering `modular transformations' of two-loop non-oriented
diagrams, in a similar way to the discrete torsion results found
in~\cite{VafaT}. We hope to return to these results in the near
future~\cite{BS2}.

\section*{Acknowledgements}

We are grateful to M. Atiyah, G. Segal and B. Totaro for useful 
discussions, and for sharing their knowledge of \Ktheory{} with us.
We are greatful to Stefan Theisen for comments on the manuscript.
We would like to thank the organisers of the Corfu Summer Institute on
Elementary Particle Physics where this work begun.

\appendix
\section{Clifford algebras and $K^{p,q}$}
\label{appendix:Clifford}

\subsubsection*{Review of Clifford algebras}

Our notation is based on~\cite{KaroubiSeminar,KaroubiPhD} but without
the category--theoretic language, see also~\cite{VBthesis}. For
completeness we review it here:
\begin{definition}
  The Clifford algebra $\CliffR{p,q}$ is the real algebra generated by
  $\gamma_1, \dots, \gamma_{p+q}$ subject to the relations
  \begin{equation}
    \label{eq:cliffordrelations}
    \begin{displayarray}{cl}
      \gamma_i \gamma_j = - \gamma_j \gamma_i &  
        \forall~ i\not=j \\
      \gamma_i^2=-1 & \forall~ i \in \{1,\dots, p\} \\
      \gamma_i^2=+1 & \forall~ i \in \{p+1,\dots, p+q\} \\
    \end{displayarray}
  \end{equation}
\end{definition}
The Clifford algebras enjoy the following well--known properties:
\begin{subequations}
\begin{gather}
  \label{eq:cliffordformula}
  \CliffR{p+n,q+n} \simeq \Mat_{2^n}\big( \CliffR{p,q} \big)
  \\
  \CliffR{p+8,q} \simeq  \CliffR{p,q+8} 
  \simeq \Mat_{16}\big(\CliffR{p,q}\big)
\end{gather}
\end{subequations}
So there are only finitely many cases to determine, the complete list
is in table~\ref{tab:clifford} (see e.g.~\cite{AtiyahBottShapiro}). 
\begin{table}[!htbp]
  \centering
  \begin{tabular}{c|cc}
    $n$ & $\CliffR{n,0}$ & $\CliffR{0,n}$ \\
    \hline
    $0$ & $\R$ & $\R{}$ \\
    $1$ & $\C$ & $\R{}\oplus \R{}$ \\
    $2$ & $\Hbb{}$ & $\Mat_2(\R{})$ \\
    $3$ & $\Hbb{}\oplus \Hbb{}$ & $\Mat_2(\C{})$ \\
    $4$ & $\Mat_2(\Hbb{})$ & $\Mat_2(\Hbb{})$ \\
    $5$ & 
    $\Mat_4(\C{})$ & 
    $\Mat_2(\Hbb{})\oplus \Mat_2(\Hbb{})$ \\
    $6$ & 
    $\Mat_8(\R{})$ & 
    $\Mat_4(\Hbb{})$ \\
    $7$ & 
    $\Mat_8(\R{})\oplus \Mat_8(\R{})$ & 
    $\Mat_8(\C{})$ 
  \end{tabular}
  \caption{List of Clifford algebras}
  \label{tab:clifford}
\end{table}
Note that the notation also reflects the multiplication in the
Clifford algebra; For example $\CliffR{0,1}$ is the algebra of pairs
$(x_1,x_2)\in \R{}\oplus \R{}$ with componentwise
multiplication. Especially $(1,0) (0,1) = (0,0)$.

Now a $\CliffR{p,q}$ vector bundle is an ordinary vector bundle $E$
with an action of the Clifford algebra, that is a map
$\rho:\CliffR{p,q}\to \End(E)$. With other words the Clifford algebra
acts on the fibres of $E$ via $\rho(\gamma_i):E_x\to E_x$.

You can add $\CliffR{p,q}$ bundles in the usual way, so by the usual
Grothendiek construction we get \Ktheory{} for bundles with
$\CliffR{p,q}$ action, denoted $KO^{(p,q)}(X)$. But those are not very
interesting groups: A real bundle with $\R{}$, $\C{}$ or $\mathbb{H}$
action is simply a real, complex or quaternionic bundle and the
semigroups of bundles and semigroups of bundles with some matrix
algebra action are isomorphic\footnote{This can be seen as follows: By
  the $\Mat_n(\R{})$ action one can decompose a bundle $E$ in the sum
  of $n$ isomorphic bundles $E=\oplus_{i=1}^n E^i$. With the
  correspondence $E\leftrightarrow E^1$ we can identify the
  semigroups.}. Furthermore if the Clifford algebra is the sum of two
orthogonal pieces (like $\R{}\oplus \R{}$) we can use the action
$\rho(1,0)$, $\rho(0,1)$ of the projectors $(1,0)$ and $(0,1)$ to
decompose the bundle into the sum of two independent bundles. So from
the last column of table~\ref{tab:clifford} we can simply read of the
\Kgroups{} in table~\ref{tab:cliffK}.
\begin{table}[htbp]
  \centering
  \begin{tabular}{ll}
    $KO^{(0,0)}(X) = KO(X)$ & 
    $KO^{(0,1)}(X) = KO(X)\oplus KO(X)$ \\
    $KO^{(0,2)}(X) = KO(X)$ & 
    $KO^{(0,3)}(X) = K(X)$ \\
    $KO^{(0,4)}(X) = KSp(X)$ & 
    $KO^{(0,5)}(X) = KSp(X)\oplus KSp(X)$ \\
    $KO^{(0,6)}(X) = KSp(X)$ & 
    $KO^{(0,7)}(X) = K(X)$ \\
  \end{tabular}
  \caption{List of the $\CliffR{0,q}$ \Kgroups}
  \label{tab:cliffK}
\end{table}

More interesting are the groups $KO^{p,q}(X)$ which fit into the long
exact sequence associated to the ``restriction of scalars''
$r:\CliffR{p,q+1}\to\CliffR{p,q}$:
\begin{equation}
  \label{eq:Kpqsequence}
  KO^{(p,q+1)}(X\times \R{}) \stackrel{r}{\to}
  KO^{(p,q)}(X\times \R{}) \to
  KO^{p,q}(X) \to  
  KO^{(p,q+1)}(X) \stackrel{r}{\to}
  KO^{(p,q)}(X)
\end{equation}
So $KO^{p,q}(X)$ is represented by formal differences of
$\CliffR{p,q+1}$ vector bundles that are the same if considered as
$\CliffR{p,q}$ vector bundles. Another way to think about that is the
following: A $\CliffR{p,q+1}$ vector bundle structure on a given
$\CliffR{p,q}$ vector bundle $E$ is equivalent to a gradation on $E$:
\begin{definition}
  \label{def:gradation}
  A gradation on a $\CliffR{p,q}$ vector bundle $(E,\rho)\in
  \mathop{\rm Vect}\nolimits_\R^{p,q}$ is an involution $\eta \in
  \End(E)$ such that $\eta^2 = 1$ and $\eta \rho(\gamma_i) = -
  \rho(\gamma_i) \eta ~\forall\, i\in \{1,\dots, p+q\}$
\end{definition}
The group $KO^{p,q}(X)$ is then the group generated by triples
$(E,\eta_1,\eta_2)$ subject to the relations
\begin{itemize}
\item $(E,\eta_1,\eta_2) + (F, \xi_1, \xi_2) = 
       (E\oplus F, \eta_1 \oplus \xi_1, \eta_2 \oplus \xi_2)$
\item $(E,\eta_1, \eta_2)=0$ if $\eta_1$ is homotopic to $\eta_2$
      within the gradations of $E$.
\end{itemize}
From these properties one can deduce the following:
\begin{itemize}
\item $(E,\eta_1, \eta_2) + (E, \eta_2,\eta_1) = 0$
\item $E\simeq E', \eta_1\simeq \eta_1', \eta_2\simeq \eta_2'
  ~\Rightarrow~ (E,\eta_1,\eta_2) = (E',\eta_1',\eta_2')
  \in KO^{p,q}(X)$ \\
  i.e. $KO^{p,q}(X)$ depends only on the isomorphism classes of bundle
  and gradations.
\item $(E,\eta_1,\eta_2) + (E,\eta_2, \eta_3) = (E,\eta_1,\eta_3)$
\item Every element of $KO^{p,q}(X)$ can be represented by a single
  triple.
\end{itemize}
One can recover ordinary $KO$-theory from $KO^{p,q}(X)$ via the following: 
\begin{equation}
  \label{eq:KO00KO}
  KO^{0,0}(X) = KO(X)    
\end{equation}
To see this let $(E,\eta_1,\eta_2)\in KO^{0,0}(X)$. Then $\eta_1$,
$\eta_2$ are just involutions, they do not have to satisfy anything
else. Maybe after adding the trivial triple $(X\times
\R^{n},\mathbf{1}_n,\mathbf{1}_n)$ one can simultaneously diagonalise
both involutions and then cancel the eigenspaces with the same
eigenvalue. You are left with a difference of triples
$(E_1,-\mathbf{1}_k,\mathbf{1}_k)-(E_2,\mathbf{1}_l,-\mathbf{1}_l)$
which you map to $[E_1]-[E_2]\in KO(X)$.

\subsubsection*{Computation of $KO_{\Zop_2}$}

All the $KO_{\Zop_2}(\R^{p,q})$ groups can be determined by using the
connection between \Ktheory{} and Clifford algebras. The basic idea is
to use the following result of~\cite{KaroubiSeminar} (for notation see
appendix~\ref{appendix:Clifford})
\begin{theorem}
  \label{theorem:fundamental}
  Let $G$ be a compact Lie group, $V$ a $G$ vector space with a
  positive definite form (i.e. the generated Clifford algebra is
  $\Cliff{(V)}$ with all $\gamma_i^2=+1$) and $X$ a Real $G$
  space. Then
  \begin{equation}
    KR_G^V(X) = KR_G(X\times V)
  \end{equation}
\end{theorem}
This reduces the calculation to one where the base space is a point, and
then use a trick to absorb the $\Zop_2$ action in the Clifford algebra.

Of course we want to compute $KO$ and not $KR$, so we want the real
version of the above theorem. So let the Real involution act trivially
on every space (that is $G$, $X$ and $V$), then 
\begin{equation}
  KO_G^V(X) = KO_G(X\times V)
\end{equation}
The correspondence between $KO \leftrightarrow KR$ classes is as usual by
complexification ($\rightarrow$) resp. taking the subbundle invariant
under the complex conjugation ($\leftarrow$). Note that for the
$V$--action to be well--defined on the real subbundle we need that the
complex conjugation acts trivially on Clifford algebra, i.e. on $V$.

So let $V=\R^{p,0}$ and let it generate $\Cliff{(V)}=\CliffR{0,p}$. We
then get
\begin{equation}
  KO_{\Zop_2}(\R^{p,q}) = KO_{\Zop_2}(\R^{0,q}\times V) = 
  KO_{\Zop_2}^V (\R^{0,q})
\end{equation}
Now we have to reinterpret $KO_{\Zop_2}^V(X)$ for some
$\Zop_2$--invariant space $X$. Its elements are tuples
$(E,g,\rho;\eta_1,\eta_2)$ where
\begin{itemize}
\item $E$ is a real vector bundle over the base space $X$.
\item $g:E\to E$ is an involution that acts trivially on the base
  (i.e. $g\in \End(E)$).
\item $g$ acts also on the Clifford algebra via $g:V\to V, v\mapsto -v$.
\item An action of the Clifford algebra $\rho:\CliffR{0,p}\to \End(E)$ 
\item Two gradations $\eta_1$, $\eta_2$.
\end{itemize}
This data is equivalent to the following:
\begin{itemize}
\item A real vector bundle $E$ over the base space $X$.
\item An action of the Clifford algebra $\tilde{\rho}:\CliffR{0,p+1}\to
  \End(E)$ defined by
  \begin{equation}
    \tilde{\rho}(\gamma_i) = \left\{
      \begin{array}{cl}
        \rho(\gamma_i) & \forall~i\in\{1,\dots,p\} \\
        g & i=p+1 \\
      \end{array}
    \right.
  \end{equation}
\item $\tilde{\eta}_1, \tilde{\eta}_2 \in \End(E)$ that commute with
  the $\CliffR{0,p+1}$ action defined by
  \begin{equation}
    \tilde{\eta}_i = g \eta_i
  \end{equation}
\end{itemize}
Since the $\tilde{\eta}_i$ {\em commute} with the Clifford algebra
action this is not $KO^{0,p+1}(X)$; The gradations and the Clifford
algebra can rather be treated independently.

However we can think of $\tilde{\eta}_i$ as two gradations of some
$\CliffR{0,0}=\R$ action on $(E,\rho)$. But then by the analog of
equation~\refeq{eq:KO00KO} for ``bundles with $\CliffR{0,p+1}$ action''
instead of ordinary bundles we find that
\begin{equation}
  KO_{\Zop_2}^V (X) = KO^{(0,p+1)}(X)
\end{equation}

\section{Understanding $H^2(\Zop_2\times \Zop_2,\Ut)$}
\label{appendix:explicitlocalcohomology}

Following (appendix~\ref{appendix:generalgroup}) we have
determined $H^2\big(\Zop_2\times\Zop_2,\Ut\big) = \Zop_2\times\Zop_2$,
so there are four different twists by which $KR_{\Zop_2\times\Zop_2}$ can
be twisted. One of them is just
untwisted $KR_{\Zop_2\times\Zop_2}$, and one of them is
$KR_{\Zop_2\times\Zop_2}^{[D_8]}$ of
section~\ref{sec:DefinitionKR}. We discuss the remaining two twists in some
detail presently. This appendix is designed to familiarise the reader
with group cohomology with coefficients in $\Ut$.

The $2$-torsion part of $H^2\big(\Zop_2\times\Zop_2,\Ut\big)$
comes from $H^2(\Zop_2\times\Zop_2,\Zop_2)$ via the $\Zop_2\to \Ut\to
\Ut$ coefficient long exact sequence. The advantage of this
description is twofold: First there is no complex conjugation action
on $\Zop_2$ and second determining $H^2(\Zop_2\times\Zop_2,\Zop_2)$ is
a finite combinatorial problem. Elementary calculation shows that
\begin{equation}
  H^2(\Zop_2\times\Zop_2,\Zop_2) = 
  \Zop_2 \times \Zop_2 \times \Zop_2
\end{equation}
corresponds to the extensions 
\begin{equation}
  0 \to \Zop_2 \to  \quad
  D_8 ~\text{or}~ 
  Q_8 ~\text{or}~ 
  \Zop_2\times \Zop_4 ~\text{or}~ 
  \Zop_2^3  
  \quad \to
  \Zop_2 \times \Zop_2 \to 0
\end{equation}
Let us denote the linear generator with $g$ and the anti-linear one
with $\tau$, so that
\begin{equation}
\label{eq:D8Q8gtaupresent}
\begin{aligned}    
  D_8 =& \Big\{ 
  \sigma, \tau, g
  \,\Big| \,
  g \tau = \sigma \tau g,~
  g^2=\tau^2=\sigma^2=1
  \Big\}
  \\
  Q_8 =& \Big\{ 
  \sigma, \tau, g
  \,\Big| \,
  g \tau = \sigma \tau g,~
  g^2=\tau^2=\sigma, \sigma^2=1
  \Big\}
\end{aligned}
\end{equation}
where the projection to $\Zop_2\times \Zop_2$ is always by putting
$\sigma=1$. To determine the corresponding group $2$-cocycle we choose
sections $s(\gamma_1)=\tau$, $s(\gamma_2)=g$, and
$s(\gamma_1\gamma_2)=\tau g$. For example
\begin{equation}
  c_{D_8}(-,-) = 
  \begin{bmatrix}
    c(\tau,\tau)   & c(g,\tau)   & c(\tau g,\tau)   \\
    c(\tau,g)      & c(g,g)      & c(\tau g,g)      \\
    c(\tau,\tau g) & c(g,\tau g) & c(\tau g,\tau g) \\
  \end{bmatrix}
  = 
  \begin{bmatrix}
    + & - & - \\ + & + & + \\ + & - & -
  \end{bmatrix}
\end{equation}
There are altogether 16 closed $2$-cocycles and one coboundary
\begin{equation}
  c(\tau)=c(g)=-1,~c(\tau g)=+1 
  \quad \Rightarrow \quad
  dc(-,-) = 
  \left[\begin{smallmatrix}
    + & - & - \\ - & + & - \\ - & - & +
  \end{smallmatrix}\right]
\end{equation}
so the quotient consists of the $8$ cohomology classes $\Zop_2\times
\Zop_2 \times \Zop_2$.

Now we are really interested in their image in
$H^2\big(\Zop_2\times\Zop_2,\Ut\big)$, for that we have to mod out the
additional coboundary 
\begin{equation}
  c(\tau)=c(g)=i,~c(\tau g)=+1 
  \quad \Rightarrow \quad
  dc(-,-) = 
  \left[\begin{smallmatrix}
    + & - & - \\ + & - & - \\ + & + & +
  \end{smallmatrix}\right]
\end{equation}
Therefore the $4$ classes are 
\begin{subequations}
\begin{align}
  c_{\Zop_2^3}(-,-) =& 
  \left[\begin{smallmatrix}
    + & + & + \\ + & + & + \\ + & + & +
  \end{smallmatrix}\right]
  \simeq
  \left[\begin{smallmatrix}
    + & - & - \\ - & + & - \\ - & - & +
  \end{smallmatrix}\right]
  \simeq
  \left[\begin{smallmatrix}
    + & - & - \\ + & - & - \\ + & + & +
  \end{smallmatrix}\right]
  \simeq
  \left[\begin{smallmatrix}
    + & + & + \\ - & - & + \\ - & - & +
  \end{smallmatrix}\right]
  \\
  c_{D_8}(-,-) =& 
  \left[\begin{smallmatrix}
    + & - & - \\ + & + & + \\ + & - & -
  \end{smallmatrix}\right]
  \simeq
  \left[\begin{smallmatrix}
    + & + & + \\ - & + & - \\ - & + & -
  \end{smallmatrix}\right]
  \simeq
  \left[\begin{smallmatrix}
    + & + & + \\ + & - & - \\ + & - & -
  \end{smallmatrix}\right]
  \simeq
  \left[\begin{smallmatrix}
    + & - & - \\ - & - & + \\ - & + & -
  \end{smallmatrix}\right]
  \\
  c_{Q_8}(-,-) =& 
  \left[\begin{smallmatrix}
    - & - & + \\ + & - & - \\ - & + & -
  \end{smallmatrix}\right]
  \simeq
  \left[\begin{smallmatrix}
    - & + & - \\ - & - & + \\ + & - & -
  \end{smallmatrix}\right]
  \simeq
  \left[\begin{smallmatrix}
    - & + & - \\ + & + & + \\ - & + & -
  \end{smallmatrix}\right]
  \simeq
  \left[\begin{smallmatrix}
    - & - & + \\ - & + & - \\ + & - & -
  \end{smallmatrix}\right]
  \\
  c_{\Zop_2\times \Zop_4}(-,-) =& 
  \left[\begin{smallmatrix}
    - & + & - \\ + & - & - \\ - & - & +
  \end{smallmatrix}\right]
  \simeq
  \left[\begin{smallmatrix}
    - & - & + \\ - & - & + \\ + & + & +
  \end{smallmatrix}\right]
  \simeq
  \left[\begin{smallmatrix}
    - & - & + \\ + & + & + \\ - & - & +
  \end{smallmatrix}\right]
  \simeq
  \left[\begin{smallmatrix}
    - & + & - \\ - & + & - \\ + & + & +
  \end{smallmatrix}\right]
\end{align}
\end{subequations}
The new twist classes corresponding to the $2$-cocycles $c_{Q_8}$,
$c_{\Zop_2\times \Zop_4}$ have $c(\tau,\tau)=-1$, so the corresponding
projective representation of $\Zop_2\times\Zop_2$ is one with
$\tau^2=-1$.

\section{The D-brane spectrum and \Kgroups{} for $\Omega\times\I_4$
orientifolds}
\label{appendix:GPorientifold}

In~\cite{QS} the D-brane spectrum of the hyper and tensor multiplet models
was computed using BCFT techniques. It is possible to generalise these
results to the two $\Omega\times\I_4$ $Sp$ orientifolds. The four
theories differ from one another by the overall choice of sign in front of
the O9- and O5-plane crosscaps.
\begin{equation}
  \begin{array}{|c||c|c|}
    \hline \strut 
    \mbox{Theory} & \quad O9 \quad & \quad O5 \quad
    \\ \hline \strut
    So \mbox{ tensor / BZDP} & \quad - \quad &  \quad + \quad \\
    So \mbox{ hyper / GP} & \quad - \quad &  \quad - \quad \\
    Sp \mbox{ tensor } & \quad + \quad &  \quad - \quad \\
    Sp \mbox{ hyper } & \quad + \quad &  \quad + \quad 
    \\\hline
  \end{array}
\end{equation}
We note in particular that the $Sp$ tensor model is T-dual (performing
T-duality along all four internal directions) to the
BZDP model. Since the $Sp$ model comes from an orbifold of a
non-supersymmetric theory (Type I with $Sp$ gauge group) this puts
in doubt the possibility of the BZDP model being 
supersymmetric\footnote{We would like to thank C. Angelantonj for 
discussions on this point.}. Like the GP model the $Sp$ hyper model
is T-dual to itself.

It is then straightforward to repeat the computation of~\cite{QS} to
obtain the D-brane spectrum of the two $Sp$ theories. We list below
the D-brane spectrum of all four models.\footnote{Entries in
  equation~\refeq{eq:Dbranespec} are of the form
  $(r_1,\cdots,r_m;s_1,\cdots,s_n)$ to indicate that all
  D$(r_i,s_j)$-branes are allowed.}

\begin{equation}\label{eq:Dbranespec}
\!\!\!\!\!\!\!\!\!\!\!\!\!\!\!  \begin{array}{|c||c|c|c|c|c|}
    \hline \strut 
    \mbox{Theory} & \Zop\oplus\Zop & \mbox{BPS } \Zop & \mbox{non-BPS } \Zop 
& \Zop_2\oplus\Zop_2 & \Zop_2 
    \\ \hline \strut
SO\mbox{ tensor / BZDP} & (1,5;0,4) & (1,3;2) & (1,5;1,2,3) 
& (-1,0;0),(3,4;4) & (-1,0;1),(3,4,3)\\
SO\mbox{ hyper / GP} & (-1,3;2) & (1,5;0,4) & (-1,3;0,1,3,4) & (5,2) 
& (5;1,3 \\
Sp \mbox{ tensor } & (1,5;0,4) & (-1,3;2) & (1,5;1,2,3) 
& (-1,0;4),(3,4;0) & (-1,0;3),(3,4;1) \\
Sp \mbox{ hyper } & (-1,3;2) & (1,5;0,4) & (-1,3;0,1,3,4) &
(1,2;2) & (1,2;1,3)
    \\\hline
  \end{array}
\end{equation}

One can compute the various twisted KR theories for $R^{0,q}$ using
equation~\refeq{eq:KD8decompose}. We only have to determine the Real
irreducible representations of the groups. For $Q_8$ we find (again 
$\delta$ denotes complex conjugation)
\begin{equation}
  \begin{array}{|c||c|c|c|c|}
    \hline \strut 
    g & 
    \quad 1 \quad & \quad \sigma \quad & 
    \quad g \quad & \quad \tau \quad 
    \\ \hline \strut
    r_1(g) & 
    \begin{pmatrix} 1 \end{pmatrix} &
    \begin{pmatrix} 1 \end{pmatrix} &
    \begin{pmatrix} 1 \end{pmatrix} &
    \begin{pmatrix} 1 \end{pmatrix}\circ \delta
    \\\hline
    r_2(g) & 
    \begin{pmatrix} 1 \end{pmatrix} &
    \begin{pmatrix} 1 \end{pmatrix} &
    \begin{pmatrix} -1 \end{pmatrix} &
    \begin{pmatrix} 1 \end{pmatrix}\circ \delta
    \\\hline
    r_3(g) & 
    \begin{pmatrix}  1 & 0 \\ 0 &  1 \end{pmatrix} &
    \begin{pmatrix} -1 & 0 \\ 0 & -1 \end{pmatrix} &
    \begin{pmatrix}  i & 0 \\  0 &  i \end{pmatrix} &
    \begin{pmatrix}  0 & 1 \\ -1 &  0 \end{pmatrix} \circ \delta
    \\\hline
    r_4(g) & 
    \begin{pmatrix}  1 & 0 \\ 0 &  1 \end{pmatrix} &
    \begin{pmatrix} -1 & 0 \\ 0 & -1 \end{pmatrix} &
    \begin{pmatrix} -i & 0 \\  0 & -i \end{pmatrix} &
    \begin{pmatrix}  0 & 1 \\ -1 &  0 \end{pmatrix} \circ \delta
    \\\hline
  \end{array}
\end{equation}
One can easily check that the commuting field for $r_3$, $r_4$ is 
$\Hbb{}$, therefore by equation~\refeq{eq:KRGdecompose}
\begin{equation}
  KR_{\Zop_2\times \Zop_2}^{[Q_8],i}(\text{pt}) = 
  KSp^i(\text{pt}) \oplus KSp^i(\text{pt}) = 
  KSp_{\,\Zop_2}^i(\text{pt}) \,.
\end{equation}
Explicitly we have for $R^{0,q}$ with trivial group action
\begin{equation}
  \begin{array}{|c||c|c|c|c|c|c|c|c|}
    \hline \strut 
    q & 
    \quad 0 \quad & \quad 1 \quad & \quad 2 \quad & 
    \quad 3 \quad & \quad 4 \quad & \quad 5 \quad &
    \quad 6 \quad & \quad 7 \quad
    \\ \hline \strut
    KSp_{\Zop}(\R^{0,q}) & 
    \Zop\oplus\Zop & 0 & 0 & 0 
& \Zop\oplus\Zop & \Zop_2\oplus\Zop_2 & \Zop_2\oplus\Zop_2 & 0
    \\\hline
  \end{array}
\end{equation}
which agrees with the spectrum of D$(5-i,4)$-branes in
equation~\refeq{eq:Dbranespec}.  Finally, take for the group defining the
remaining twist $[D_8 Q_8]$
\begin{equation}    
  D_8 Q_8 = \Big\{ 
  \sigma, \tau, g
  \,\Big| \,  g \tau = \tau g,~
  g^2=\tau^2=\sigma, \sigma^2=1
  \Big\}
\end{equation}

Its Real irreducible representations are 
\begin{equation}
  \begin{array}{|c||c|c|c|c|}
    \hline \strut 
    g & 
    \quad 1 \quad & \quad \sigma \quad & 
    \quad g \quad & \quad \tau \quad 
    \\ \hline \strut
    r_1(g) & 
    \begin{pmatrix} 1 \end{pmatrix} &
    \begin{pmatrix} 1 \end{pmatrix} &
    \begin{pmatrix} 1 \end{pmatrix} &
    \begin{pmatrix} 1 \end{pmatrix}\circ \tau 
    \\\hline
    r_2(g) & 
    \begin{pmatrix} 1 \end{pmatrix} &
    \begin{pmatrix} 1 \end{pmatrix} &
    \begin{pmatrix} -1 \end{pmatrix} &
    \begin{pmatrix} 1 \end{pmatrix}\circ \tau 
    \\\hline
    r_3(g) & 
    \begin{pmatrix}  1 & 0 \\ 0 &  1 \end{pmatrix} &
    \begin{pmatrix} -1 & 0 \\ 0 & -1 \end{pmatrix} &
    \begin{pmatrix}  0 & 1 \\ -1 &  0 \end{pmatrix} &
    \begin{pmatrix}  0 & 1 \\ -1 &  0 \end{pmatrix} \circ \tau
    \\\hline
  \end{array}
\end{equation}
The only higher-dimensional Real representation $r_3$ has commuting
field $\C$, and therefore
\begin{equation}
  KR_{\Zop_2\times \Zop_2}^{[D_8 Q_8],i}(\text{pt}) = 
  K^i(\text{pt})
\end{equation}
This matches with the spectrum of D$(5-i,4)$-branes in 
equation~\refeq{eq:Dbranespec}.

\section{Cohomology for general abelian groups}
\label{appendix:generalgroup}

In this section we want to compute the cohomology group
$H^2\big(G,\Ut\big)$ for all finite abelian groups
\begin{equation}
  G = \Zop_{2r} \times
  \Big( \mathop{\times}_{i=1}^{n} \Zop_{k_i} \Big) \times
  \Big( \mathop{\times}_{j=1}^{m} \Zop_{\ell_j} \Big) 
  \qquad k_i~\text{even},~\ell_j~\text{odd}
\end{equation}
where the augmentation $\epsilon:G\to \Zop_2$ is $-1$ on the generator of the
first factor and $+1$ on the other generators. By redefinition of the
generators we can always assume that only one generator acts anti-linearly.

Now it is technically easier to use the $\Zt\to \Rt\to \Ut$ coefficient
long exact sequence (for finite groups $H^i(G,\Rt)=0$)
\begin{equation}
  \label{eq:exponentialsequence}
  0 \to 
  H^2\left(G,\Ut\right) 
  \stackrel{\sim}{\longrightarrow}
  H^3\left(G,\Zt\right)
  \to  0
\end{equation}
and actually compute $H^3\big(G,\Zt\big)$. Then the computation
naturally splits into two steps:
\begin{enumerate}
\item Compute the group cohomology for general cyclic groups
\item Put the cohomology groups of the factors together via the
  K\"unneth theorem
\end{enumerate}
The former is standard~\cite{Brown}
\begin{theorem}
  Let $G=\Zop_k$ be any cyclic group and $M$ an arbitrary $\Zop[G]$
  module. Then the cohomology groups are
  \begin{equation}
    H^i(G,M) = 
    \begin{cases}
      \ker N & i~\text{odd} \\
      \coker N = M^G/NM & i>0~\text{even} \\
      M^G & i=0
    \end{cases}
  \end{equation}
  where\footnote{The invariants and coinvariants of $M$ are defined as
    \begin{equation}
      M^G =\, \{ m\in M|~g(m)=m \} 
      \quad \text{and} \quad
      M_G =\, M \big/ \{g(m)-m|~ g\in G, m\in M\}
    \end{equation}
%%     \begin{equation}
%%       \begin{split}
%%         M^G =&\, \{ m\in M|~g(m)=m \} \\
%%         M_G =&\, M \big/ \{g(m)-m|~ g\in G, m\in M\}
%%       \end{split}
%%     \end{equation}
}  $N:M_G\to M^G$ is the norm map $N(m) = \sum_{i=1}^k g^i(m)$.
\end{theorem}
In particular we have
%% \begin{equation}
%%   H^i(\Zop_{2r},\Zt) = 
%%   \begin{cases}
%%     0           \\
%%     \Zop_2      \\
%%     0           \\
%%     \Zop_2      \\
%%     0           
%%   \end{cases}
%% %  \begin{cases}
%% %    0           & i=4 \\
%% %    \Zop_2      & i=3 \\
%% %    0           & i=2 \\
%% %    \Zop_2      & i=1 \\
%% %    0           & i=0
%% %  \end{cases}
%%   H^i(\Zop_{k_j},\Zop) = 
%%   \begin{cases}
%%     \Zop_{k_j}  \\
%%     0           \\
%%     \Zop_{k_j}  \\
%%     0           \\
%%     \Zop        \\
%%   \end{cases}
%%   H^i(\Zop_{k_j},\Zop) = 
%%   \begin{cases}
%%     \Zop_{k_j}  & i=4 \\
%%     0           & i=3 \\
%%     \Zop_{k_j}  & i=2 \\
%%     0           & i=1 \\
%%     \Zop        & i=0
%%   \end{cases}
%% \end{equation}
\begin{equation}
  \label{eq:groupcohomologycyclic}
  \begin{tabular}{l|ccccc}
    & $i=0$ & $i=1$ & $i=2$ & $i=3$ & $i=4$  \\ \hline
    $H^i\big(\Zop_{2r},\Zt\big)$ 
      \raisebox{1ex}{\strut}
      & $0$ & $\Zop_2$ & $0$ & $\Zop_2$ & $0$ \\
    $H^i\big(\Zop_{k_j},\Zop\big)$ 
      & $\Zop$ & $0$ & $\Zop_{k_j}$ & $0$ & $\Zop_{k_j}$ \\
    $H^i\big(\Zop_{\ell_j},\Zop\big)$ 
      & $\Zop$ & $0$ & $\Zop_{\ell_j}$ & $0$ & $\Zop_{\ell_j}$ \\
  \end{tabular}
\end{equation}
To assemble the cohomology groups of $G$ from the factors we then need
the
\begin{theorem}[K\"unneth theorem]
  Let $G_1$, $G_2$ be groups (such that resolutions are finitely
  generated, e.g. finite groups). Furthermore let $M_1$ be a
  $\Zop[G_1]$ module and $M_2$ a $\Zop[G_2]$ module such that either
  $M_1$ or $M_2$ is $\Zop$-free. Then there is a split exact
  sequence
  \begin{multline}
    0 \longrightarrow
    \bigoplus_{p+q=i} \Big( H^p(G_1,M_1) \otimes H^q(G_2,M_2) \Big)
%    \longrightarrow \\ \longrightarrow
    \longrightarrow
    H^i(G_1\times G_2, M_1\otimes M_2)
    \longrightarrow \\ \longrightarrow
    \bigoplus_{p+q=i+1}
      \Tor \Big( H^p(G_1,M_1), H^q(G_2,M_2) \Big) 
    \longrightarrow 0
  \end{multline}
\end{theorem}
Since in our case the cohomology groups $H^\ast(\Zop_{2r},\Zt)$ are
only $2$-torsion and the above theorem implies that the
sequence splits we see~\footnote{Recall that
  $\Tor(\Zop_n,\Zop_m)=\Zop_{\gcd(n,m)}$,
  $\Tor(\Zop_n,\Zop)=0=\Tor(\Zop,\Zop_n)$.} that the cohomology
of the general abelian group $G$ is only $2$-torsion. This matches
the physical expectation that only $\pm1\in \Uone$ twists are allowed 
(see section~\ref{sec:PhysInt}).

Moreover the odd order factors in $G$ do not contribute: viewed as
a $\Zop$-module equation we have 
$\Zop_2 \otimes \Zop_{\ell}=0=\Tor(\Zop_2,\Zop_{\ell})$ for $\ell$ odd
we see that
\begin{equation}
  H^\ast\left(\Zop_{2r} \times \Zop_{\ell}, \Zt\right) = 
  H^\ast\left(\Zop_{2r}, \Zt\right) 
  \qquad \ell~\text{odd}
\end{equation}
so we only have to consider the cyclic subgroups of even order. For
those the relevant cohomology groups can be summarised as follows:
\begin{theorem}
  For $\displaystyle G_n \eqdef \Zop_{2r} \oplus 
  \Big( \mathop{\oplus}_{i=1}^{n} \Zop_{k_i} \Big)$ we have
  \begin{equation}
    \begin{tabular}{l|cccc}
      & $i=0$ & $i=1$ & $i=2$ & $i=3$  \\ \hline
      $H^i\big(G_n,\Zt\big)$ 
        \raisebox{1ex}{\strut}
        & $0$ & $\Zop_2$ & $\Zop_2^n$ & $\Zop_2^{1+n(n+1)/2}$
    \end{tabular}
  \end{equation}
\end{theorem}
\begin{proof}
  Induction: It is correct for $n=0$
  (equation~\refeq{eq:groupcohomologycyclic}). Then by the K\"unneth theorem
%%   \begin{align}
%%     H^0\big(G_{n+1},\Zt\big) =& \Zop \otimes H^0(G_n,\Zt) = 0    
%%     \\
%%     H^1\big(G_{n+1},\Zt\big) =& \Zop \otimes H^1(G_n,\Zt) = \Zop_2
%%     \\    
%%     H^2\big(G_{n+1},\Zt\big) =& H^2(G_n,\Zt) \oplus 
%%       \Tor\Big(H^1(G_n,\Zt),H^2(\Zop_{k_{n+1}},\Zop)\Big) = 
%%       \Zop_2 \oplus \Zop_2^n
%%     \\
%%     H^3\big(G_{n+1},\Zt\big) =& H^3(G_n,\Zt) \oplus 
%%       \Big( H^1(G_n,\Zt) \otimes H^2(\Zop_{k_{n+1}},\Zop) \Big) 
%%       \oplus \notag \\ & \hspace{4cm} 
%%       \oplus \Tor\Big(H^2(G_n,\Zt),H^2(\Zop_{k_{n+1}},\Zop)\Big)=
%%       \notag \\ =&
%%       \Zop_2^{1+n(n+1)/2} \oplus \Zop_2 \oplus \Zop_2^n = 
%%       \Zop_2^{1+(n+1)(n+2)/2}
%%   \end{align}
  \begin{subequations}
  \begin{align}
    H^0\big(G_{n+1},\Zt\big) =& 0    
    \\
    H^1\big(G_{n+1},\Zt\big) =& \Zop_2
    \\    
    H^2\big(G_{n+1},\Zt\big) =& \Zop_2 \oplus \Zop_2^n = \Zop_2^{n+1}
    \\
    H^3\big(G_{n+1},\Zt\big) =& 
      \Zop_2^{1+n(n+1)/2} \oplus \Zop_2 \oplus \Zop_2^n = 
      \Zop_2^{1+(n+1)(n+2)/2}
  \end{align}
  \end{subequations}
\end{proof}
Putting everything together we have learnt that
\begin{equation}
  H^2\left(
    \Zop_{2r} \oplus
    \big( \mathop{\oplus}_{i=1}^{n} \Zop_{k_i} \big) \oplus
    \big( \mathop{\oplus}_{j=1}^{m} \Zop_{\ell_j} \big) 
  , \Ut \right)
  = \Zop_2^{1+n(n+1)/2}
  \qquad k_i~\text{even},~\ell_j~\text{odd} \,.
\end{equation}


\begin{thebibliography}{99}

\bibitem{QS}
N.~Quiroz and B.~Stefa\'nski, jr,
{\em Dirichlet branes on orientifolds},
[hep-th/0110041].

\bibitem{VBthesis}
V.~Braun,
{\em K--Theory and Exceptional Holonomy in String Theory},
PhD~thesis, to appear.

\bibitem{PolRR}
J.~Polchinski, 
{\em Dirichlet-Branes and Ramond-Ramond Charges},
Phys.\ Rev.\ Lett.\  {\bf 75} (1995) 4724, 
[hep-th/9510017].
%%CITATION = HEP-TH 9510017;%%

\bibitem{Snbps} 
A.~Sen, 
{\em Stable non-BPS states in string theory}, 
JHEP {\bf 9806} (1998) 007,
[hep-th/9803194];
%%CITATION = HEP-TH 9803194;%%

A.~Sen, 
{\em Stable non-BPS bound states of BPS D-branes}, 
JHEP {\bf 9808} (1998) 010,
[hep-th/9805019];
%%CITATION = HEP-TH 9805019;%%

A.~Sen, 
{\em Tachyon condensation on the brane antibrane system}, 
JHEP {\bf 9808} (1998) 012,
[hep-th/9805170];
%%CITATION = HEP-TH 9805170;%%

A.~Sen, 
{\em SO(32) spinors of type I and other solitons on brane-antibrane pair}, 
JHEP {\bf 9809} (1998) 023, 
[hep-th/9808141];
%%CITATION = HEP-TH 9808141;%%

A.~Sen, 
{\em Type I D-particle and its interactions}, 
JHEP {\bf 9810} (1998) 021,
[hep-th/9809111].
%%CITATION = HEP-TH 9809111;%%

A.~Sen, 
{\em BPS D-branes on non-supersymmetric cycles}, 
JHEP {\bf 9812} (1998) 021
[hep-th/9812031].
%%CITATION = HEP-TH 9812031;%%

\bibitem{BG} 
O.~Bergman and M.~R.~Gaberdiel, 
{\em Stable non-BPS D-particles}, 
Phys.\ Lett.\ B {\bf 441} (1998) 133, 
[hep-th/9806155].
%%CITATION = HEP-TH 9806155;%%

\bibitem{WittK} 
E.~Witten, 
{\em D-branes and \Ktheory}, 
JHEP {\bf 9812} (1998) 019, 
[hep-th/9810188].
%%CITATION = HEP-TH 9810188;%%

\bibitem{Frau} 
M.~Frau, L.~Gallot, A.~Lerda and P.~Strigazzi,
{\em Stable non-BPS D-branes in type I string theory}, 
Nucl.\ Phys.\ B {\bf 564} (2000) 60,
[hep-th/9903123].
%%CITATION = HEP-TH 9903123;%%

\bibitem{GS} 
M.~R.~Gaberdiel and B.~Stefa\'nski, jr., 
{\em Dirichlet branes on orbifolds}, 
Nucl.\ Phys.\ B {\bf 578} (2000) 58, 
[hep-th/9910109].
%%CITATION = HEP-TH 9910109;%%

\bibitem{Srednicki} 
M.~Srednicki, 
{\em IIB or not IIB}, 
JHEP {\bf 9808} (1998) 005,
[hep-th/9807138].
%%CITATION = HEP-TH 9807138;%%

\bibitem{MM} 
R.~Minasian and G.~Moore, 
{\em \Ktheory{} and Ramond-Ramond charge},
JHEP {\bf 9711} (1997) 002, 
[hep-th/9710230].
%%CITATION = HEP-TH 9710230;%%

\bibitem{BianchiSagnotti} 
M.~Bianchi and A.~Sagnotti,
{\em Twist Symmetry And Open String Wilson Lines},
Nucl.\ Phys.\ B {\bf 361} (1991) 519.
%%CITATION = NUPHA,B361,519;%%

\bibitem{AS}
C.~Angelantonj and A.~Sagnotti, {\em Open strings},
arXiv:hep-th/0204089.
%%CITATION = HEP-TH 0204089;%%

\bibitem{gomis}
J.~Gomis, {\em D-branes on orbifolds with discrete torsion and 
topological  obstruction},
JHEP {\bf 0005} (2000) 006
[arXiv:hep-th/0001200].
%%CITATION = HEP-TH 0001200;%%

\bibitem{GP} 
E.~G.~Gimon and J.~Polchinski, 
{\em Consistency Conditions for Orientifolds and D-Manifolds}, 
Phys.\ Rev.\ D {\bf 54} (1996) 1667, 
[hep-th/9601038].
%%CITATION = HEP-TH 9601038;%%

\bibitem{DP}
A.~Dabholkar and J.~Park, 
{\em A note on orientifolds and F-theory},
Phys.\ Lett.\ B {\bf 394} (1997) 302, 
[hep-th/9607041].
%%CITATION = HEP-TH 9607041;%%

\bibitem{BZ} 
J.~D.~Blum and A.~Zaffaroni, 
{\em An orientifold from F theory}, 
Phys.\ Lett.\ B {\bf 387} (1996) 71,
[hep-th/9607019].
%%CITATION = HEP-TH 9607019;%%

\bibitem{VafaT} 
C.~Vafa, 
{\em Modular Invariance And Discrete Torsion On Orbifolds},
Nucl.\ Phys.\ B {\bf 273} (1986) 592.
%%CITATION = NUPHA,B273,592;%%

\bibitem{z2torsion} 
C.~Vafa and E.~Witten, 
{\em On orbifolds with discrete torsion},
J.\ Geom.\ Phys.\  {\bf 15} (1995) 189, 
[hep-th/9409188].
%%CITATION = HEP-TH 9409188;%%

M.~R.~Douglas, 
{\em D-branes and discrete torsion}, 
[hep-th/9807235].
%%CITATION = HEP-TH 9807235;%%

M.~R.~Douglas and B.~Fiol, 
{\em D-branes and discrete torsion. II},
[hep-th/9903031].
%%CITATION = HEP-TH 9903031;%%

\bibitem{SCY} 
B.~Stefa\'nski, jr., 
{\em Dirichlet branes on a Calabi-Yau three-fold orbifold},
Nucl.\ Phys.\ B {\bf 589} (2000) 292, 
[hep-th/0005153].
%%CITATION = HEP-TH 0005153;%%

\bibitem{HorK}
P.~Horava, {\em Type IIA D-branes, K-theory, and matrix theory},
Adv.\ Theor.\ Math.\ Phys.\  {\bf 2} (1999) 1373
[arXiv:hep-th/9812135].
%%CITATION = HEP-TH 9812135;%%

\bibitem{BGH} 
O.~Bergman, E.~G.~Gimon and P.~Horava, 
{\em Brane transfer operations and T-duality of non-BPS states}, 
JHEP {\bf 9904} (1999) 010, 
[hep-th/9902160].
%%CITATION = HEP-TH 9902160;%%

\bibitem{KleinRab}
M.~Klein and R.~Rabadan, {\em Orientifolds with discrete torsion},
JHEP {\bf 0007} (2000) 040
[arXiv:hep-th/0002103].
%%CITATION = HEP-TH 0002103;%%

\bibitem{SugimotoGB}
O.~Bergman, E.~G.~Gimon and S.~Sugimoto, {\em Orientifolds, RR torsion, 
and K-theory},
JHEP {\bf 0105} (2001) 047
[arXiv:hep-th/0103183].
%%CITATION = HEP-TH 0103183;%%

\bibitem{GabCY} 
M.~R.~Gaberdiel, 
{\em Discrete torsion orbifolds and D-branes}, 
JHEP {\bf 0011} (2000) 026, 
[hep-th/0008230].
%%CITATION = HEP-TH 0008230;%%

\bibitem{KaroubiPhD}
M.~Karoubi,
{\em Alg{\`e}bres de {C}lifford et {K}--th{\'e}orie},
Ann.\ Scient.\ {\'E}c.\ Norm.\ Sup., {\bf 4(1)} (1968) 161--270.

\bibitem{KaroubiBook}
M.~Karoubi,
{\em {K}--Theory},
Springer, 1978.

\bibitem{KaroubiSeminar}
M.~Karoubi, R.~Gordon, Peter L{\"o}ffler, and M.~Zisman,
{\em S{\'e}minaire Heidelberg--Saarbr{\"u}cken--Strasbourg sur la
  {K}--Th{\'e}orie}, 
volume 136 of Lecture Notes in Mathematics, Springer, 1970.

\bibitem{AtiyahBottShapiro}
M.~F.~Atiyah, R.~Bott, and A.~Shapiro,
{\em Clifford modules},
Topology, {\bf 3(1)} (1964) 3--38.

\bibitem{Totaropriv}
B.~Totaro,
{\em Private Communications}. 

\bibitem{Atiyahpriv}
M.~F.~Atiyah,
{\em Private Communications}. 

\bibitem{Segalpriv}
G.~B.~Segal,
{\em Private Communications}.

\bibitem{AtiyahSegalEquiv}
M.~F.~Atiyah and G.~B.~Segal,
{\em Equivariant {K}-theory and completion},
J. Diff. Geom. \textbf{3} (1969) 1--18.

\bibitem{SharpeDT}
E.~R.~Sharpe,
{\em Discrete Torsion and Gerbes I},
[hep-th/9909108].

\bibitem{Gukov}
S.~Gukov,
{\em \Ktheory, Reality, and Orientifolds},
Commun.\ Math.\ Phys. {\bf 210} (2000) 621--639,
[hep-th/9901042].

\bibitem{Serre}
J.~P.~Serre, {\em Linear representations of finite groups}, 
Springer-Verlag, 1996.

\bibitem{SchomerusWZWa}
A.~Alekseev and V.~Schomerus,
%``RR charges of D2-branes in the WZW model,''
arXiv:hep-th/0007096.
%%CITATION = HEP-TH 0007096;%%

\bibitem{SchomerusWZWb}
S.~Fredenhagen and V.~Schomerus,
%``Branes on group manifolds, gluon condensates, and twisted K-theory,''
JHEP {\bf 0104} (2001) 007
[arXiv:hep-th/0012164].
%%CITATION = HEP-TH 0012164;%%

\bibitem{FreedWitten}
D.~S.~Freed, E.~Witten,
{\em Anomalies in String Theory with D-Branes},
[hep-th/9907189].
%%CITATION = HEP-TH 9907189;%%

\bibitem{BouwknegtMathai}
P.~Bouwknegt, V.~Mathai,
{\em D-branes, B-fields and twisted K-theory},
JHEP\ \textbf{0003} (2000) 007,
[hep-th/0002023].

%\cite{Kapustin:1999di}
\bibitem{Kap}
A.~Kapustin, {\em D-branes in a topologically nontrivial B-field},
Adv.\ Theor.\ Math.\ Phys.\  {\bf 4} (2000) 127
[arXiv:hep-th/9909089].
%%CITATION = HEP-TH 9909089;%%


\bibitem{RabUranga}
R.~Rabadan and A.~M.~Uranga, {\em Type IIB orientifolds without 
untwisted tadpoles, and non-BPS D-branes},
JHEP {\bf 0101} (2001) 029
[arXiv:hep-th/0009135].
%%CITATION = HEP-TH 0009135;%%

\bibitem{PolT}
J.~Polchinski, {\em Tensors from K3 orientifolds},
Phys.\ Rev.\ D {\bf 55} (1997) 6423
[arXiv:hep-th/9606165].
%%CITATION = HEP-TH 9606165;%%

\bibitem{Brown}
K.~S.~Brown, {\em Cohomology of Groups}, Springer-Verlag, 1982.

\bibitem{BS2}
V.~Braun, B.~Stefa\'nski, jr., {\em Orientifolds and K-theory II},
work in progress.


\end{thebibliography}
\end{document}